\documentclass[preprint,11pt,number,3p,times]{elsarticle}
\usepackage[latin1]{inputenc}
\usepackage[T1]{fontenc}

\bibstyle{elsarticle-num}

\usepackage[centertags]{amsmath}
\usepackage{amssymb}
\usepackage{hyperref}
\hypersetup{
    bookmarks=false,      % show bookmarks bar?
    pdfstartview={FitH},  % fits the width of the page to the window
    colorlinks=true,      % false: boxed links; true: colored links
    linkcolor=black,      % color of internal links
    citecolor=blue,       % color of links to bibliography
    filecolor=blue,       % color of file links
    urlcolor=blue,        % color of external links
    pdftitle = {The Non-SUSY Orbifolder},
    pdfauthor = {Escalante-Notario, Perez-Martinez, Ramos-Sanchez, Vaudrevange}
}

\usepackage{color}
\usepackage{epsfig}
\usepackage{varioref}
\usepackage{scrtime}
\usepackage{bbm}
\usepackage{dsfont}
\usepackage{nicefrac}
\usepackage{longtable}
\usepackage{lscape}
\usepackage{listings}

\usepackage{braket}

\usepackage[usenames,dvipsnames,table]{xcolor}
\usepackage{fancyvrb}

\usepackage[textsize=tiny,backgroundcolor=yellow]{todonotes}

\newcommand{\code}[1]{\small{\tt #1}\normalsize}
\newcommand{\orbifolder}{\code{non-SUSY\,orbifolder}}

\newcommand{\Id}{\ensuremath{\mathds{1}}}
\newcommand{\SO}[1]{\ensuremath{\mathrm{SO}(#1)}}
\newcommand{\SU}[1]{\ensuremath{\mathrm{SU}(#1)}}
\newcommand{\U}[1]{\ensuremath{\mathrm{U}(#1)}}
\newcommand{\E}[1]{\ensuremath{\mathrm{E}_{#1}}}
\newcommand{\Z}[1]{\ensuremath{\mathbbm{Z}_{#1}}} % z_N ->\Z{N}

\usepackage{listings}
\usepackage{xcolor}

\definecolor{darkgray}{rgb}{0.4,0.4,0.4}
\definecolor{purple}{rgb}{0.65, 0.12, 0.82}
\definecolor{backgroundgray}{rgb}{0.95,0.95,0.95}

\definecolor{dockerblue}{RGB}{0,122,204}
\definecolor{dockergray}{RGB}{64,64,64}
\definecolor{dockerlightgray}{RGB}{240,240,240}
\definecolor{dockergreen}{RGB}{0,150,100}

\lstdefinelanguage{Dockerfile}{
    keywords={FROM, RUN, CMD, LABEL, MAINTAINER, EXPOSE, ENV, ADD, COPY, ENTRYPOINT, VOLUME, USER, WORKDIR, ARG, ONBUILD, STOPSIGNAL, HEALTHCHECK, SHELL},
    keywordstyle=\color{blue}\bfseries,
    comment=[l]{\#},
    commentstyle=\color{gray}\itshape,
    stringstyle=\color{red},
    morestring=[b]",
    inputencoding=utf8,
    extendedchars=true,
    literate={á}{{\'a}}1 {ñ}{{\~n}}1 {ó}{{\'o}}1 {í}{{\'{\i}}}1 {é}{{\'e}}1 {ú}{{\'u}}1,
}

\definecolor{DarkGrey}{rgb}{0.1,0.1,0.1}
\definecolor{gray97}{gray}{.91}
\definecolor{gray75}{gray}{.75}
\definecolor{gray45}{gray}{.45}
\makeatother

\lstdefinestyle{consola}
{backgroundcolor=\color{gray97},
basicstyle=\small\color{black}\ttfamily,
breaklines=false,
inputencoding=utf8,
extendedchars=true,
columns=flexible,
literate={á}{{\'a}}1 {ñ}{{\~n}}1 {ó}{{\'o}}1 {í}{{\'{\i}}}1 {é}{{\'e}}1 {ú}{{\'u}}1,
}

\newcommand{\I}{\mathrm{i}}
\newcommand{\e}{\mathrm{e}}

\newcommand{\maR}{\ensuremath{\mathbbm{R}} }
\newcommand{\maZ}{\ensuremath{\mathbbm{Z}} }

\newcommand{\maT}{\ensuremath{\mathbbm{T}} }

\newcommand{\x}{\ensuremath{\times}}
\usepackage{cancel, color}

\definecolor{darkgreen}{HTML}{109930}  %
\journal{Computer Physics Communications}

\begin{document}

\begin{frontmatter}

  \title{\textbf{\boldmath The Non-SUSY Orbifolder:\unboldmath}\\
    a tool to build promising non-supersymmetric string models}

  \author[escom]{Enrique~Escalante-Notario} \ead{eescalanten@ipn.mx}
  \author[uac]{Ricardo~P\'erez-Mart\'inez} \ead{ricardo.perezmartinez@uadec.edu.mx}
  \author[unam]{Sa\'ul~Ramos-S\'anchez} \ead{ramos@fisica.unam.mx}
  \author[]{Patrick~K.S.~Vaudrevange} \ead{patrickvaudrevange@gmail.com}
  \address[escom]{Escuela Superior de C\'omputo, Instituto Polit\'ecnico Nacional, C.P. 07738, Cd.~de M\'exico, M\'exico.}
  \address[uac]{Facultad de Ciencias F\'isico-Matem\'aticas, Universidad Aut\'onoma de Coahuila, Edificio A, Unidad Camporredondo, 25000, Saltillo, Coahuila, M\'exico}
  \address[unam]{Instituto de F\'isica, Universidad Nacional Aut\'onoma de M\'exico, Cd.~de M\'exico C.P.~04510, M\'exico}

  \begin{abstract}
    We introduce the \orbifolder{}, which is a program developed in \code{C++} that computes
    the low-energy effective theory of non-supersymmetric heterotic orbifold compactifications.
    The program includes routines to compute the massless spectrum, to automatically generate large
    sets of orbifold models, to identify phenomenologically interesting models (e.g.\ models sharing features
    of the Standard Model (SM) or Grand Unified Theories (GUT)) and to analyze their vacuum-configurations.
  \end{abstract}

\end{frontmatter}

\section*{Program Summary}
\noindent{\em Program title:} \orbifolder{}\\
{\em Program obtainable
from:} {\tt https://github.com/StringsIFUNAM/nonSUSYorbifolder}\\
{\em Requests and questions
in:} {\tt https://github.com/StringsIFUNAM/nonSUSYorbifolder/issues}\\
{\em Distribution formats:}\/ .zip, .tgz, Docker Image\\
{\em Programming language:} {\code{C++}}\\
{\em Computer:}\/ Personal computer\\
{\em Operating system:}\/ Tested on Linux (Ubuntu 16.04, 18.04, 20.04, 22.04, 24.04), Mint 21, and Fedora 39\\
{\em Word size:}\/ 64 bits\\
{\em External routines:}\/ None\\
{\em Dependencies:} Boost, GSL, Readline \\
{\em Typical running time:}\/ Less than a second per model.\\
{\em Nature of problem:}\/ Calculating the low-energy spectrum of non-supersymmetric heterotic orbifold compactifications.\\
{\em Solution method:}\/ Quadratic equations on a lattice; representation theory; polynomial algebra. \\

%%%%%%%%%%%%%%%%%%% NEW SECTION %%%%%%%%%%%%%%%%%%%
\clearpage
\newpage
\section{Introduction}

In the quest to determine whether string theory offers a consistent description of Nature,
a first challenge is to reconcile it with the fundamental physics we observe.
To accomplish this, the six extra spatial dimensions in string theory must be compactified,
ensuring that the resulting 4-dimensional (4D) effective field theory encompasses the
properties of the Standard Model (SM) of particle physics. Most efforts in this direction
rely on 10-dimensional (10D) string theories that incorporate supersymmetry (SUSY). However,
since SUSY has not yet been detected, it is essential to develop tools that allow us to explore
string-theory frameworks that could give rise to non-supersymmetric (non-SUSY) 4D models
with phenomenologically viable properties.

Consistent non-SUSY 4D effective field theories can be achieved by compactifying the 10D non-SUSY heterotic string
with gauge group \SO{16}\x\SO{16}~\cite{Gross:1984dd,Dixon:1986iz,Alvarez-Gaume:1986ghj} (see e.g.~\cite{Baykara:2024tjr, Larotonda:2024thv} for newer different approaches),
which automatically avoids some frequent pathologies, such as tachyons and (local and global) anomalies, see e.g.~\cite{Basile:2023knk}.
Several approaches have been successfully pursued in this endeavor~\cite{Abel:2015oxa,Ashfaque:2015vta,Blaszczyk:2015zta,Abel:2017vos,Faraggi:2020fwg,Aoyama:2020aaw,Aoyama:2021kqa,Faraggi:2020wld,Faraggi:2020hpy,Florakis:2021bws,Faraggi:2022hut}, with symmetric, Abelian, toroidal orbifold
compactifications (heterotic orbifolds for short)~\cite{Blaszczyk:2014qoa,Perez-Martinez:2021zjj} standing out as the
simplest formalism that leads to promising results (for details on heterotic orbifolds,
see also~\cite{Dixon:1985jw,Dixon:1986jc,Bailin:1999nk,Ramos-Sanchez:2008nwx,Vaudrevange:2008sm,Ramos-Sanchez:2024keh}).
Also beyond the heterotic string, some important effort has been done to study the properties and phenomenological potential
of 4D non-SUSY string models~\cite{Sagnotti:1996qj,Angelantonj:1998gj,Blumenhagen:1999ns,Moriyama:2001ge,Gato-Rivera:2007ifz}.

As we explain to some extent in section~\ref{sec:orbis}, non-SUSY heterotic orbifolds are characterized by a large set of parameters. This includes those describing the spatial transformations mapping the six extra dimensions of the heterotic string, $\maR^6$, into a compact toroidal orbifold. Further parameters emerge from the inevitable embedding of these spatial transformations into the gauge degrees of freedom. Among all possible 6D orbifolds, there are 138 distinct geometric configurations for symmetric, Abelian toroidal orbifolds~\cite{Fischer:2012qj} that are consistent with the demand of a vanishing cosmological constant~\cite{GrootNibbelink:2017luf}. Moreover, each of these compactification spaces admits a vast number of gauge embeddings that satisfy the modular-invariance conditions~\cite{Ploger:2007iq}, which are crucial for ensuring the consistency of the constructions. This complexity explains why much of the research in this area relies on computational tools to identify phenomenologically viable compactifications and investigate their associated phenomenology.

In the case of heterotic orbifolds with SUSY, there are already some useful tools to learn more about the physics of string theory. For example, the supersymmetric \code{orbifolder}~\cite{Nilles:2011aj} has been instrumental in many of the recent developments of string phenomenology arising from heterotic orbifolds~\cite{Lebedev:2007hv,Lebedev:2008un,Goodsell:2011wn,Carballo-Perez:2016ooy,Olguin-Trejo:2018wpw,Olguin-Trejo:2019hxk}. More recently, methods based on neural networks have also been successfully implemented~\cite{Mutter:2018sra,Parr:2019bta,Parr:2020oar,Escalante-Notario:2022fik}. However, so far, no dedicated software is publicly available to study non-SUSY heterotic orbifolds.

This work introduces the \orbifolder{}, a tool developed in \code{C++} that shares some features with the supersymmetric \code{orbifolder}~\cite{Nilles:2011aj}. The \orbifolder{} is designed to construct tachyon-free, anomaly-free, and consistent non-SUSY orbifolds, which lead to non-SUSY 4D effective quantum field theories, exhibiting phenomenologically desirable properties. These include gauge symmetries and a massless matter spectrum resembling that of the SM and extensions, such as Grand Unified Theories (GUTs). In addition, the \orbifolder{} provides a set of commands for analyzing some of the phenomenological aspects of these models. For example, its import/export functionality enables users to perform new studies on previously generated model libraries, such as those of refs.~\cite{Perez-Martinez:2021zjj,Cervantes:2023wti}.

While the absence of SUSY might seem like a straightforward correction to supersymmetric codes, the \orbifolder{} addresses a series of complex challenges that are not present with SUSY. As detailed in~\cite{Blaszczyk:2014qoa}, these challenges include transitioning from supersymmetric to non-SUSY heterotic strings, eliminating tachyons and anomalies, and distinguishing between bosons and fermions in the spectra, among other critical considerations.

This work is organized as follows. In section~\ref{sec:orbis} we discuss some details of non-SUSY heterotic orbifolds in order to guide the reader through the notation of our string constructions. In section~\ref{sec:download}, we provide a brief set of instructions for the user interested in downloading our software. Then, section~\ref{sec:structure} is devoted to the main features of the basic element to work with the \orbifolder{}, \texttt{the prompt}. Sections~\ref{sec:fromzero},~\ref{sec:randommodels} and~\ref{sec:loadmodels} are meant as a fundamental guide for the user interested in creating non-SUSY models. The conclusions and outlook are presented in section~\ref{sec:conclusions}. A glossary of all commands is provided in~\ref{app:commands}, and an explanation for using the \orbifolder{} with a Docker Container is presented in~\ref{app:docker}.

%%%%%%%%%%%%%%%%%%% NEW SECTION %%%%%%%%%%%%%%%%%%%
\section{Non-SUSY heterotic orbifold compactifications}
\label{sec:orbis}

Let us briefly introduce the formalism of the non-SUSY heterotic string and orbifold compactifications in order to set the notation and conventions used in the \orbifolder{}.

The 10D non-SUSY heterotic string with gauge group $\SO{16} \x \SO{16}$ exhibits a massless spectrum that consists of 240 gauge bosons and 512 fermions. It also includes a gravitational sector built by the graviton, the Kalb-Ramond field and the dilaton. This theory is modular invariant on the 2D world-sheet and free of tachyons and anomalies~\cite{Gross:1984dd,Dixon:1986iz,Alvarez-Gaume:1986ghj,Blaszczyk:2014qoa}.

One particularly simple scheme to arrive at phenomenologically {appealing results in 4D} from the non-SUSY heterotic strings is a toroidal orbifold {compactification of six spatial dimensions.} An orbifold is defined as the quotient space ${\cal{O}} = \maR^6/S$, where $\maR^6$ describes the 6D flat space to compactify, and $S$ is the so-called space group, which can be generated for example by
\begin{itemize}
\item discrete rotations that define the point group $P = \Z{M} = \{\theta^m\,\vert\,\theta^{M} = \Id\}$, where $m = 0,1,2,\ldots,M-1$, and $\theta = \text{diag}(1,\e^{2\pi \I v^1},\e^{2\pi \I v^2},\e^{2\pi \I v^3})$ is the generator of \Z{M}. The vector $v = (0,v^1,v^2,v^3)$ is called the twist vector and encodes the rotation angles on the three 2D planes of $\maR^6$.

\item translational vectors that are elements of a 6D lattice defined as $\Lambda = \{m_\alpha e_\alpha\,\vert\, m_\alpha\,\in\,\maZ\}$, where $e_\alpha$, $\alpha = 1,2,\ldots,6$, are its basis vectors, and summation over $\alpha$ is implied. More general vectors, i.e.\ translations that are not elements of the lattice $\Lambda$, are called roto-translations. They are represented as $r_\alpha e_\alpha$, where $r_\alpha\,\notin\,\maZ$ and necessarily accompany a rotation, such that they act as e.g.\ $x\mapsto\theta\,x+r_\alpha e_\alpha$ for $x\in\maR^6$.
\end{itemize}

The point group can also be $P = \Z{M}\x\Z{N} = \{\theta^m \omega^n\,\vert\,\theta^M\omega^N=\Id\}$, where $m=0,1,2,\ldots,M-1$ and $n=0,1,2,\ldots,N-1$. In this case there are two twist vectors, $v_1$ and $v_2$, that define the generators $\theta$ and $\omega$ for $\Z{M}$ and $\Z{N}$, respectively. The components of the twist vector $v$ for a \Z{M} point group satisfy the condition $v^1 + v^2 + v^3 = 0$, and similarly each of the two twist vectors $v_1$ and $v_2$ for orbifolds with point group $\Z{M}\x\Z{N}$. The previous condition guarantees ${\mathcal{N}}=1$ SUSY in 4D when the orbifold compactification is performed on the 10D ${\cal{N}} = 1$ SUSY heterotic string\footnote{The notation ${\cal{N}} = 1$ indicates the number ${\cal{N}}$ of supersymmetry generators of the theory. Here we prefer to name the corresponding ${\cal{N}} = 0$ heterotic string just as the non-SUSY heterotic string.}. These orbifolds are called SUSY orbifolds. When there are no roto-translations the orbifold can also be defined as ${\cal{O}} = \maT^6/P$, where $\maT^6$ is a 6D torus and $P$ is the point group. The torus is defined as $\maT^6 = \maR^6/\Lambda$. For this reason, $\Lambda$ is called the torus lattice, and the corresponding orbifold is named toroidal orbifold. In general, $P$ must be a symmetry of the torus lattice $\Lambda$, and $P\,\subset\,O(6)$. The point group $P$ is a discrete finite group that can be Abelian or non-Abelian, leading to the names Abelian or non-Abelian toroidal orbifolds. In this work we focus on Abelian toroidal orbifolds. The classification of the SUSY (Abelian and non-Abelian) toroidal orbifolds was performed in reference~\cite{Fischer:2012qj}, where the authors found 138 Abelian orbifold geometries (space groups) with point groups $P = \Z{M}$ or $P =\Z{M}\x\Z{N}$ that can be used for a consistent orbifold compactification of the heterotic string and, as found in ref.~\cite{GrootNibbelink:2017luf}, that lead to a vanishing cosmological constant for the compactification of non-SUSY heterotic strings. These are the space groups that the \orbifolder{} is able to use to compactify~\cite{Blaszczyk:2014qoa,Perez-Martinez:2021zjj}.

The elements of the space group $S$ with point group $\Z{M}\x\Z{N}$ are denoted as $g=(\theta^m\omega^n, n_\alpha e_\alpha)$, where $n_\alpha$ can be an integer or {a fraction. Only in the presence of roto-translations is $n_\alpha$ fractional}. The action of $S$ on $x\,\in\,\maR^6$ is defined as $x\mapsto x':=g\,x = \theta^m\omega^n x + n_\alpha e_\alpha$. When the point group acts non-trivially on the 6D space coordinates and $x_f' = x_f$, {$x_f$ is a fixed point in the orbifold. Consequently,} $x_f=(\Id-\theta^m\omega^n)^{-1} n_\alpha e_\alpha$
if the matrix $(\Id-\theta^m\omega^n)$ is non-singular, otherwise so-called fixed tori appear. The space-group element associated with a fixed point is called its constructing element. A conjugate class of the space group is built by conjugate elements.\footnote{Let $g,g',h$ be elements of a group $G$, then the element $g$ is conjugate of $g'$, with respect to $h$, if $hgh^{-1}=g'$.} There are infinitely many fixed points. However, it is possible to identify a finite number of inequivalent fixed points since all space-group elements in a conjugate class are associated with the same (inequivalent) fixed point in the quotient space of the orbifold ${\cal{O}}$. Then, any element of a conjugate class can be taken as the constructing element of the corresponding fixed point.

The 10D non-SUSY heterotic string theory $\SO{16}\x\SO{16}$ {can be constructed from the 10D SUSY heterotic string \E8\x\E8 by {letting a \Z2 point group act on the six extra dimensions of the SUSY string. This group, denoted as \Z{2W}, is generated by the Witten twist $\beta$, such that $\Z{2W} = \{\beta^k\,\vert\,\beta^2 = \Id\}$, $k=0,1$, and acts freely (i.e.\ without fixed points) on the 6D spatial coordinates, but it has a non-trivial action on the fermions of the theory~\cite{Rohm:1983aq,Gross:1984dd,Dixon:1986iz,Alvarez-Gaume:1986ghj, Blaszczyk:2014qoa}.} The Witten twist can be encoded in the Witten twist vector $v_0 = (0,1,1,1)$, whose role is similar to the previously defined twist vectors.}
Taking into account the freely-acting group \Z{2W}, we have that elements of the space group are denoted as $g = (\beta^k\theta^m\omega^n, n_\alpha e_\alpha)$, {with full point group $\Z{K}\x\Z{M}\x\Z{N}=\Z{2W}\x\Z{M}\x\Z{N}$, where $k = 0,1$, $m = 0,1,2,\ldots,M-1$, and $n = 0,1,2,\ldots\,N-1$. It is understood that only the point group \Z{M}\x\Z{N} (or \Z{M}) is responsible of the orbifold compactification of the non-SUSY heterotic string whereas \Z{2W} is required to build the non-SUSY heterotic string.}

Modular invariance of the heterotic string is a strong consistency constraint that guarantees the absence of anomalies~\cite{Gross:1984dd,Dixon:1986iz}. It requires the embedding of the space group $S$ into the {gauge degrees of freedom $X^I$, $I = 1,2,\ldots,16$, as a so-called gauge twisting group $G$.} The action of $G$ on the gauge coordinates is given by $X^I\mapsto X'^I:=\tilde{g} X^I = X^I + \pi(k V^I_0 + m V^I_1 + n V^I_2 + n_\alpha W^I_\alpha)$, where $\tilde{g} = (k V^I_0 + m V^I_1 + n V^I_2, n_\alpha W^I_\alpha)\,\in\,G$ represents the gauge embedding of $g=(\beta^k\theta^m\omega^n, n_\alpha e_\alpha)\,\in\,S$. The 16D vectors $V_0$, $V_1$ and $V_2$ are called shift vectors and the six 16D vectors $W_\alpha$ are named Wilson lines~\cite{Ibanez:1986tp}. In particular, the vector $V_0 := (1,0^7;1,0^7)$ is the Witten shift vector and corresponds to the embedding of the Witten twist vector $v_0$ into the gauge degrees of freedom of the heterotic string~\cite{Dixon:1986iz}. {Both $v_0$ and $V_0$ are fixed and essential to construct the 10D non-SUSY heterotic string from the supersymmetric theory. In contrast, the shift vectors $V_1$ and $V_2$ and the six Wilson lines $W_\alpha$ can be dialed to build the best candidate models to fit observations, as long as they satisfy a number of consistency conditions. Explicitly, shift vectors and Wilson lines need to comply with modular invariance constraints associated with the anomaly freedom of the resulting 4D effective field theories~\cite{Ploger:2007iq,Blaszczyk:2014qoa}. These conditions read
\begin{equation}\begin{aligned}
\label{eq:modInv}
M(V_1\cdot V_1-v_1\cdot v_1) &= 0\,\bmod\,2\,,\qquad&
N(V_2\cdot V_2-v_2\cdot v_2) &= 0\,\bmod\,2\,,\\
N(V_1\cdot V_2-v_1\cdot v_2) &= 0\,\bmod\,2\,,&
V_0\cdot V_i = 0\,\bmod\,1 &= V_0\cdot W_\alpha\,,\\
M_\alpha\, (V_i\cdot W_\alpha)&=0\,\bmod\,2\,,&
M_\alpha\, (W_\alpha\cdot W_\alpha) &= 0\,\bmod\,2\,,\\
Q_{\alpha\beta}\, (W_\alpha\cdot W_\beta) &= 0\,\bmod\,2\,, &(\alpha\neq\beta\,,\quad \text{no sum}&\text{ implied})
\end{aligned}\end{equation}
where $M,N$ and $M_\alpha$ are the orders of $V_1,V_2$ and $W_\alpha$, respectively,} i.e.\ $M\,V_1, N\,V_2$ and $M_\alpha W_\alpha$ (no sum in $\alpha$) are elements of the 16D root lattice $\Gamma_{16}$ of the gauge group \SO{16}\x\SO{16}. {Further, $Q_{\alpha\beta}:=\text{gcd}(M_\alpha,M_\beta)$.} %, whose origin is due to the compactification of the 16 internal coordinates $X^I, I=1,2,\ldots,16$, on a 16D torus such that the modular invariance of the theory and the absence of SUSY impose that the 16D torus lattice must be the $\SO{16}\times\SO{16}$ root lattice.
{The simplest} way to construct a consistent gauge twisting group is choosing standard embedding, {i.e.\ letting the non-trivial structures of shift and twist vectors coincide and taking vanishing Wilson lines.}\footnote{{Standard embedding of a \Z{M}\x\Z{N} orbifold with twist vector $v_i = (0,v^1_i,v^2_i,v^3_i)$ consists in taking $V_i=(V^1_i,V^2_i,V^3_i,0^5,0^8)$ with $V_i^a=v_i^a$, for $i=1,2,a=1,2,3$.}}

{The massless spectrum is typically divided in states arising from} untwisted and twisted sectors. The closed-string boundary conditions for strings are modified {in the orbifold}, such that the 6D spatial coordinates to be compactified on a 6D orbifold {must satisfy} $X(\tau,\sigma+\pi) = g\,X(\tau,\sigma)$, where $\tau$ and $\sigma$ are the time- and space-like coordinates of the 2D world-sheet, and $g=(\beta^k\theta^m\omega^n, n_\alpha e_\alpha)\,\in\,S$. The untwisted sector comprises strings satisfying the boundary conditions with the constructing element $g=(\Id,0)\,\in\,S$, and the twisted sectors {are related to} $g=(\beta^k\theta^m\omega^n,n_\alpha e_\alpha)\,\in\,S$, with non-zero $k, m$ or $n$. Untwisted strings are original closed strings and twisted strings close only up to the non-trivial action of the space group, i.e. they are attached to fixed points characterized by their constructing elements $g\,\in\,S$. The 4D massless states come from the tensor products of left-moving states with right-moving states that satisfy the level-matching condition, $M^2_L = 0 = M^2_R$, and that are invariant under the space group and the gauge twisting group. The massless untwisted sector consists of the original massless states of the \SO{16}\x\SO{16} non-SUSY heterotic string that are compatible with the symmetry transformations of the orbifold group. {In particular, since not all of the original gauge bosons are invariant under the orbifold, only a subgroup $G_{4D}\subset\SO{16}\x\SO{16}$ is realized as the 4D gauge group. Also, only the components of a 4D graviton remain invariant from the original 10D graviton.}
For twisted strings, the {properties of} left- and right-moving massless states are obtained from the conditions
\begin{equation}
  M^2_L = \frac{(p+V_g)^2}{2}+\tilde{N}-1+\delta_c = 0 \qquad\text{and}\qquad
  M^2_R = \frac{(q+v_g)^2}{2}-\frac{1}{2}+\delta_c=0\,,
  \label{eq:M0}
\end{equation}
where $p$ and $q$ are the momenta of left- and right-moving strings{, which are weights of} the $\SO{16}\x\SO{16}$ and $\SO{8}$ weight lattices, respectively. The number operator {$\tilde{N}$ counts the number of} left-moving oscillators, and {$\delta_c=\delta_c(v_1,v_2)$ is a computable shift in the zero-point energy produced by the orbifolds twists $v_1,v_2$}. The vector $V_g := kV_0 + mV_1 + nV_2 + n_\alpha W_\alpha$ is {frequently referred to as} local shift vector, and $v_g := kv_0 + mv_1 + nv_2$ is the local twist vector. The twisted massless states are represented by $\tilde{\alpha}\ket{p+V_g}\otimes\ket{q+v_g}$ where $\tilde{\alpha}$ are possible oscillators excitations for the left-moving states. The shifted momenta are defined as $p_{sh} := p + V_g$ and $q_{sh} := q + v_g$.  The gauge momenta $p_{sh}$ and $p$ describe {weights of the gauge representations built by the corresponding} twisted and untwisted massless states under $G_{4D}$, respectively. Finally, since twisted strings are attached to fixed points that are characterized by their constructing elements, we have that space group elements $g = (\beta^k\theta^m\omega^n, n_\alpha e_\alpha)$ identify the localization of twisted strings or twisted sectors as $(k,m,n;n_\alpha)$.

%We close this section by establishing the conventions and notations used in the \orbifolder{}. We
{As a closing remark on the notation used here, we shall frequently label an orbifold model by its compactification point group $\Z{M}\times \Z{N}$ or \Z{M}, omitting the freely-acting twist $\Z{2W}$. This is evident in the names of the geometry (and model) files that the program reads to build the models. Internally though, the full point group $\Z{K}\x\Z{M}\x\Z{N}$ is used. This is evident in the localization of the states, which is labeled by $(k,m,n;n_\alpha)$ as mentioned earlier.}  %We also have that all commands and outputs in the \orbifolder{} follow this convention, except some commands that display details of the fields in the \code{spectrum} directory, where the sector and localization of twisted strings given by $(k,m,n;n_\alpha)$ is necessary for a complete description of the scalar and fermion massless fields. Finally, the information contained in the model and geometry files corresponds to the orbifold with point group $\Z{K}\x\Z{M}\x\Z{N}$, indicating that the freely acting group $\Z{K}=\Z{2W}$ is used in the \orbifolder{} to construct the 10D non-SUSY heterotic string $\SO{16}\x \SO{16}$ from the 10D SUSY heterotic string $\E8\x\E8$.

\subsection{SM-like models}
\label{subs:smlike}

{One of the prime purposes of the \orbifolder{} is to build phenomenologically viable orbifold compactifications. So, we aim at constructing models based on compactifications of the non-SUSY heterotic string that replicate the observed features of the SM, while also offering new elements that could address outstanding problems in particle physics and cosmology.  We call such models SM-like.} An orbifold model {is considered to be SM-like if it displays} the following properties:
\begin{itemize}
\item The 4D gauge group is $G_{4D} = {\cal{G}}_{\text{SM}} \x G_{\text{hidden}} \x \U1^n$, where ${\cal{G}}_{\text{SM}} = \SU3_c\x \SU2_L\x \U1_Y$ is the SM gauge group, and the hypercharge is non-anomalous and {its generator $Y$ is} compatible with \SU5 grand unification, {for which we adopt the normalization $Y\cdot Y = \nicefrac56$}. $G_{\text{hidden}}$ is a non-Abelian gauge group, usually consisting of products of $\SU{N}$ {and/or $\SO{2N}$} group factors. It is called hidden because almost none of the SM fields are charged under this group. The number $n$ of Abelian \U1 gauge symmetries is such that the rank of $G_{4D}$ is 16.

\item The 4D massless spectrum consists of the SM particles plus some exotic particles, i.e.\ particles that {are not included in the SM. The spectrum must contain} the three generations of chiral fermions, including three right-handed neutrinos, and at least one Higgs doublet. {All extra} fermions {must build vector-like pairs} with respect to $\cal{G}_{\text{SM}}$. {An admissible SM-like model can exhibit} exotic scalars and a number of SM-singlet fermions and scalars.
\end{itemize}

%%%%%%%%%%%%%%%%%%% NEW SECTION %%%%%%%%%%%%%%%%%%%
\section{Download and Installation}
\label{sec:download}

{Let us now provide the basics that allow an interested user to install the \orbifolder{} in their computer.}
In this section, we will assume that the operating system where the \orbifolder{} is to be installed
is a GNU/Linux distribution, {such as Ubuntu 22.04 or superior}. For installation on other
operating systems, such as Windows 10 or MacOS 15.1, please refer to our~\ref{app:docker}.

{The source code of the \orbifolder{} is stored and managed via GitHub.} This provides numerous advantages both for those interested in contributing
improvements to the source code and for users who are only interested in running the software. GitHub
not only acts as a centralized repository that facilitates access to the source code but also
offers a suite of collaborative tools for software development, such as a Git-based version
control system, which allows detailed tracking of changes to the code, making it easier to identify
and fix errors. Furthermore, GitHub encourages user participation through issue reporting,
improvement suggestions, and even contributions of new features via pull requests. This ensures
that the project evolves continuously and maintains high quality. Finally, downloading and
installing the \orbifolder{} becomes a simple and well-documented process, significantly
improving the user experience.

To obtain the source code for the \orbifolder{}, the most efficient option is to clone the
repository directly from GitHub using Git. This method allows you to download an exact copy of
the project, preserving all the version history and existing branches. To do so, simply ensure
that Git is installed on your system, which can be easily done by opening a terminal and running
the command
\begin{lstlisting}[style=consola,numbers=none]
$ sudo apt-get install git
\end{lstlisting}

To clone the repository containing the source code, {one must run the command}
\begin{lstlisting}[style=consola,numbers=none]
$ git clone https://github.com/StringsIFUNAM/nonSUSYorbifolder.git
\end{lstlisting}
By cloning the repository, not only is the complete source code obtained, but also the organized
structure of the project, facilitating the compilation and development process. Additionally, this
option allows easy synchronization of future changes made in the repository with the local system
using the \texttt{git pull} command, ensuring that you always work with the most up-to-date version
of the program. This is especially useful for developers planning to contribute to the project or
users who require the latest improvements and bug fixes.

As a direct alternative to obtain the source code for the \orbifolder{}, a compressed folder
containing all the necessary files to compile and run the program is available at
\begin{center}
    \href{http://stringpheno.fisica.unam.mx/nonSUSYorbifolder}{http://stringpheno.fisica.unam.mx/nonSUSYorbifolder}
\end{center}
This option is ideal for users who prefer to avoid using tools like Git or who are simply looking
for a quick and straightforward way to access the project. By downloading the compressed file,
the user receives a specific copy of the latest version of the source code, ready to be
decompressed and used. This alternative eliminates the need to manage a local repository or
learn version control commands, making the process more accessible for beginners or those
unfamiliar with GitHub. However, it is important to note that, unlike cloning the repository,
this option does not include the ability to track automatic updates or collaborate directly on
the project's development. Once the folder is downloaded, it must be decompressed, and the user
should navigate in it.

{Independently of how the code is obtained,}
once inside the folder containing the \orbifolder{} source code, the next step is to
install the necessary libraries to compile and run the program correctly. These libraries include
Boost C++, GNU Scientific Library (GSL), and GNU Readline, which provide essential functionalities
such as advanced mathematical computations and support for command-line interfaces. On
Ubuntu-based systems, these can be easily installed via the \code{apt} package manager by executing
in a Linux terminal the commands
\begin{lstlisting}[style=consola,numbers=none]
$ sudo apt-get update
$ sudo apt-get install libgsl0-dev libboost-math-dev libreadline-dev
\end{lstlisting}
This process ensures that all dependencies are available on the system before proceeding with
the source-code compilation.

With the dependencies installed, the next step is to compile the \orbifolder{} source code.
For this, a \texttt{C++} compiler such as \texttt{g++} is required, which is usually available on Linux-based
systems. Within the source-code folder, the compilation process is automatized by using a
Makefile, which contains the necessary instructions to compile the program. It suffices to execute
\begin{lstlisting}[style=consola,numbers=none]
$ ./configure
$ make
$ make install
\end{lstlisting}
These commands will configure and locate the \texttt{Makefile} in the current directory, executing the defined steps to compile the source code, such as linking the previously installed libraries and generating the final executable file. During this process, it is important to pay attention at any error or warning messages, as they may indicate issues with missing dependencies or incorrect system configurations. Once the compilation has been successfully completed, an executable file is generated in the same directory, ready to be executed.

%%%%%%%%%%%%%%%%%%% NEW SECTION %%%%%%%%%%%%%%%%%%%
\section{How to run the program}
\label{sec:structure}

{The \orbifolder{} can be easily run in a terminal window from the installation folder. Depending on the user's goal, the program offers three methods to execute it. By directly calling the executable}
\begin{lstlisting}[style=consola,numbers=none]
$ ./nonSUSYorbifolder
\end{lstlisting}
{the \orbifolder{} offers a ``blank canvas'' to create new orbifold models as explained in detail in section~\ref{sec:fromzero}, with no prior data uploaded. One may instead be interested in analyzing some previously created model. This can be achieved by executing the program with an argument:}
\begin{lstlisting}[style=consola,numbers=none]
$ ./nonSUSYorbifolder model_definitions.txt
\end{lstlisting}
{where \code{model\_definitions.txt} is a text file containing one or more sets of input data that define previously found orbifold models. Examples of such a file can be found in the \code{Models} directory, located in the installation folder of the \orbifolder{}. We provide one sample model for each of the 138 orbifold geometries admitted by the software. The previous two methods offer a Linux-style command line, called {\it the prompt}, where instructions can be typed to study the details of the constructions.

Finally, one can run the program providing the argument {\it script} and a file name,}
\begin{lstlisting}[style=consola,numbers=none]
$ ./nonSUSYorbifolder script set_of_commands.txt
\end{lstlisting}
{where \code{set\_of\_commands.txt} is a text file containing a set of admissible commands. The commands included in the file are run, one by one, and their output is stored automatically for later analysis.

In the remainder of this section, we provide some details about the use of the prompt and the script.}

\subsection{The prompt}
\label{subs:prompt}

The prompt is a Linux-style command line that allows us to interact with the \orbifolder{}. It is structured in directories where commands are defined to develop different tasks. As mentioned earlier, to start working with the {prompt, we can run the command \code{./nonSUSYorbifolder}. This command by itself initiates the prompt with no prior data. In order to better exemplify the use of the prompt, we run the \orbifolder{} loading a \Z3 orbifold model by executing the instruction}
\begin{lstlisting}[style=consola,numbers=none]
$ ./nonSUSYorbifolder Models/ZN_models/modelZ3_1_1.txt
\end{lstlisting}
The \orbifolder{} initializes and sets the prompt in the main directory denoted as \code{>}. \code{Z3\_1\_1} labels the uploaded orbifold model.
The welcoming message reads
\begin{lstlisting}[style=consola,numbers=none]
########################################################################################
#  Non-SUSY Orbifolder                                                                           #
#  Version: 1.0                                                                                  #
#  by E. Escalante-Notario, R. Perez-Martinez, S. Ramos-Sanchez and P.K.S. Vaudrevange           #
########################################################################################

Load orbifolds from file "Models/ZN_models/modelZ3_1_1.txt".
Orbifold "Z3_1_1" loaded.

>
\end{lstlisting}
{The same result is obtained by running \code{./nonSUSYorbifolder} in the terminal window, and then executing} the command \code{load orbifold(Models/ZN\_models/modelZ3\_1\_1.txt)} to load the \Z3 orbifold model.

To see the contents of the main directory type
\begin{lstlisting}[style=consola,numbers=none]
> dir
\end{lstlisting}
Then, a list of commands is displayed along with the option to enter the orbifold directory where the orbifold model was loaded. The name of the orbifold directory corresponds to the orbifold label, which in the current example is \code{Z3\_1\_1}. To enter this directory, type
\begin{lstlisting}[style=consola,numbers=none]
> cd Z3_1_1
\end{lstlisting}
and then
\begin{lstlisting}[style=consola,numbers=none]
/Z3_1_1> dir
\end{lstlisting}
to display {the general available commands along with the five orbifold-model} subdirectories
\begin{equation}
  \label{code:dirs}
  \code{model},\; \code{gauge\,group},\; \code{spectrum},\; \code{vev-config}\; \text{and}\; \code{vev-config/labels}\,.
\end{equation}
They are common to all orbifold models, and they contain several {instructions that allow us to study the models}. For example, in the \code{model} directory the user can print and change input data for the orbifold model geometry, see~\ref{subs:modeldir}. In the \code{gauge group} directory one can print details of the gauge group, see~\ref{subs:ggdir}. In the \code{spectrum} directory is possible to print several details of the orbifold model spectrum, see~\ref{subs:spectdir}. In the \code{vev-config} directory, the user can define and analyze the vev-configuration for the orbifold models and select the observable sector, see~\ref{subs:vdir}. In the \code{vev-config/labels} directory one can work with the labels for the fields in the spectrum, see~\ref{subs:labeldir}. Also, in the main directory of the \orbifolder{} the user can create, load, rename and delete orbifolds, see~\ref{subs:maindir}. It is also the directory where one can access the orbifold model directories that were created or loaded.

Let us consider a brief example to show the use of some commands. To print the gauge group of this orbifold model, i.e.\ the 4D gauge group that results from the compactification of the non-SUSY heterotic $\SO{16}\times\SO{16}$ string on the \Z3 orbifold, enter the \code{gauge group} directory
\begin{lstlisting}[style=consola,numbers=none]
/Z3_1_1> cd gauge group
\end{lstlisting}
and execute the command
\begin{lstlisting}[style=consola,numbers=none]
Z3_1_1/gauge group> print gauge group
\end{lstlisting}
Then, the gauge group $\SO{10} \x \SU{3} \x \SO{16} \x \U1$ is shown. The simple roots\footnote{They are 16D vectors that help to specify the 4D gauge group of the orbifold model, among other group theoretical properties~\cite{Ramos-Sanchez:2008nwx, Vaudrevange:2008sm}.} can be obtained with the command
\begin{lstlisting}[style=consola,numbers=none]
/Z3_1_1/gauge group> print simple roots
\end{lstlisting}
In the \code{spectrum} directory the 4D massless spectrum of this orbifold model can be printed. For this purpose, go back to the orbifold model directory using
\begin{lstlisting}[style=consola,numbers=none]
/Z3_1_1/gauge group> cd ..
\end{lstlisting}
and enter the \code{spectrum} directory
\begin{lstlisting}[style=consola,numbers=none]
/Z3_1_1> cd spectrum
\end{lstlisting}
then, use the command
\begin{lstlisting}[style=consola,numbers=none]
/Z3_1_1/spectrum> print summary
\end{lstlisting}
The massless spectrum for scalar and fermion fields is now printed according to the representations and charges under the  4D gauge group $\SO{10} \x \SU{3} \x \SO{16} \x \U1$. The shortcuts \code{m,gg,s,v,l} allow the user to enter directly to the directories in~\eqref{code:dirs}. For example, to go to the \code{model} directory, just type \code{m}. It is also possible to exit the specific model directory and return to the main directory by typing \texttt{cd \raisebox{0.29ex}{\texttildelow}}.

In sections~\ref{sec:fromzero},~\ref{sec:randommodels} and~\ref{sec:loadmodels} we provide more detailed examples to create different types of orbifold models and to explore some of their properties. We also show a sample input and output for the \Z3 orbifold with the standard embedding in the additional material~\cite[\S Complementary notes, Table 3]{website:2025}. A glossary of all commands in the \orbifolder{} is presented in~\ref{app:commands}.

\subsubsection{Helping utilities}
\label{subs:help-man}

{The \orbifolder{} prompt offers two utilities to help us know the available commands in each directory. The command \code{dir} or, equivalently, \code{help} displays the instructions available in the current directory as well as some details on their arguments. In some cases, \code{help} accepts some arguments to provide further information about one particular command. For example, in the main directory \code{>}, one can type \code{help create random} to learn more about how to initiate the random creation of consistent orbifold models with a given geometry. Other arguments for \code{help} in the main directory are \code{help system commands} and \code{help processes}, which display the instructions available in all directories, such as general settings for the \orbifolder{} kernel and instructions to deal with processes that are run in the background. Another useful example is \code{help short cuts}, which can be executed in an orbifold-model directory (once a model has been created) and displays a practical set of 1-2 key instructions (\code{m,gg,s,v,l}) to quickly access any of the subdirectories within an orbifold-model folder.

In addition, in every directory one can use the standard \code{man} command to access a manual for specific commands (see~\ref{sub:man} for more details). All available manuals in the current directory are displayed by typing \code{man} with no arguments. The manuals contain detailed explanations of the commands as well as useful examples.
}

\subsection{The script}
\label{subs:script}

The script refers to a list of commands that the user can write in a file. These commands are executed by the \orbifolder{} and the results are written in another file that is automatically created. Let us explain briefly how it works. First, write a list of commands in a file and save it in the directory where the \orbifolder{} is installed. Suppose the name of the file is \code{commands.txt}. Next, run the \orbifolder{} by using the instruction \code{./nonSUSYorbifolder script commands.txt}. Then, the \orbifolder{} executes the commands and shows a message indicating that the results were written in a file named \code{result\_commands.txt}, which is automatically created and saved in the directory where the \orbifolder{} is installed. Now, the user can see the results of the commands by opening the file \code{result\_commands.txt}.

As an example, let us consider the commands used in section~\ref{subs:prompt},
\begin{lstlisting}[style=consola,numbers=none]
load orbifold(Models/ZN_models/modelZ3_1_1.txt)
cd Z3_1_1
dir
cd gauge group
print gauge group
print simple roots
cd ..
cd spectrum
print summary
\end{lstlisting}
and save the file with the name \code{commands.txt} (any file name is admissible). Now, we run the program with the arguments
\begin{lstlisting}[style=consola,numbers=none]
$ ./nonSUSYorbifolder script commands.txt
\end{lstlisting}
The commands are then executed and their outputs are saved in a file named \code{result\_commands.txt}. {One can use the possibility to print the output} in a different file. For this purpose, instead of writing e.g.\ \code{print summary}, we enter \code{print summary to file(Filename)}, where \code{Filename} can be a more convenient file.

\subsection{Files defining an orbifold model}
\label{subs:geomandmodelfiles}

The \orbifolder{} uses two files to define an orbifold model: i) The geometry file provides the space group as the geometric information of the orbifold, and ii) the model file gives the shift vectors and Wilson lines that act on the gauge sector of the heterotic string. Examples are given in the additional material~\cite[\S Complementary notes]{website:2025}. In those notes, Table 1 contains {a thorough description of a sample geometry file, and Table 2, a sample model} file for the \Z3 orbifold with standard embedding. {Let us discuss here some useful features.}

%The \orbifolder{} uses geometry and model files to define and create orbifold models.  provide the space group information like the twist and translation generators, the identical Wilson lines and their order, the 6D torus lattice, the constructing elements and their centralizers elements.  The model files present the shift vectors and Wilson lines that define an orbifold model. There are examples of these files for the \Z{3} orbifold in the additional material~\cite[\S Complementary notes]{website:2025}: Table 1 shows the geometry file and Table 2 the model file. Next, we give a brief description of these files.

\subsubsection{The geometry file}
\label{subs:geomf}

The geometry file provides:
\begin{itemize}
  \item The point group. It is \Z{K}\x\Z{M}\x\Z{N} or \Z{K}\x\Z{M}, where \Z{K}=\Z{2W} is used for the construction of the 10D non-SUSY heterotic string and \Z{M}\x\Z{N} or \Z{M} is used for the compactification of this theory. In the geometry file, the orders of the three group factors in  \Z{K}\x\Z{M}\x\Z{N}, i.e.\ $2, M, N$, are listed. For example, for the \Z{3} orbifold model (internally \Z{2W}\x\Z{3}) the numbers 2, 3, 1 are shown, where 1 indicates the absence of the \Z{N} factor group in this case.

  \item The lattice label. It indicates the name or label of the 6D torus lattice of the corresponding orbifold geometry. For example, for the \Z{3} (i.e.\ \Z{2W}\x\Z{3}) orbifold there exists only one distinct lattice, which is denoted by Z3\_1.

  \item The twist vectors. The point group \Z{2W}\x \Z{M} \x \Z{N} has twist vectors $v_0, v_1$ and $v_2$, respectively. For example, the twists vectors for the \Z{3} orbifold (i.e.\ \Z{2W}\x\Z{3}) are $v_0 = (0,1,1,1)$ and $v_1 = (0,\frac{1}{3},\frac{1}{3},-\frac{2}{3})$. The Witten twist vector $v_0$ is fixed in all orbifold models.

  \item The twist space group generators. For each twist vector, the twist space group generators list the rotational generators of the space group, hence, including the case of roto-translations. Then, a space group generator $g = (\beta^k\theta^m\omega^n,n_\alpha e_\alpha)\,\in\,S$ is represented by nine numbers: three integers $k$, $m$, and $n$, and six {rational numbers} $n_\alpha$.

  \item The shift space group generators. It indicates the six translational generators $(\Id,e_\alpha)$, $\alpha = 1,2,\ldots,6$ of the 6D torus lattice, represented by nine numbers: $k=m=n=0$ and six integers $n_\alpha\in\{0, 1\}$.

  \item The 6D torus lattice. It presents the six 6D basis vectors, $e_1,e_2,\ldots,e_6$, for the 6D torus lattice. The lattice is factorizable if it can be written as $\maT^6 = \maT^2\x\maT^2\x\maT^2$, otherwise it is non-factorizable.

  \item The identical Wilson lines and their orders. According to the space group properties, a Wilson line has finite order and some Wilson lines can be constrained to be identical. For example, for the \Z{3} orbifold the order of the six Wilson lines is 3, and the identical Wilson lines are $W_1 = W_2$, $W_3 = W_4$ and $W_5 = W_6$.

  \item The constructing elements. The space group elements that correspond to inequivalent fixed points in the orbifold are called space group constructing elements. They are presented in this part of the geometry files.

  \item The centralizer elements. The centralizer of a constructing element denotes the set of all space group elements that commute with the constructing element. For each centralizer, a set of generators is listed in this part of the geometry files.

\end{itemize}

\subsubsection{The model file}
\label{subs:modelf}

The model file provides:
\begin{itemize}
  \item The label of the orbifold model. For example, for the \Z3 (i.e.\ \Z{2W}\x\Z{3}) orbifold model the label is \code{Z3\_1\_1}. The label \code{Z3\_1\_1} is also the name of the directory where the orbifold model is stored when it is loaded.

  \item The space group. Indicates the filename ({including} directory) where the corresponding orbifold geometry file is located. The standard location of the geometry files is the folder named \code{Geometry}. For the \Z3 orbifold, the geometry file is named \code{Geometry\_Z3\_1\_1.txt} (see section~\ref{subs:notenames} for an explanation of the names of these files).

  \item The 16D lattice. The \SO{16}\x\SO{16} root lattice of the non-SUSY heterotic string {arises in the \orbifolder{}} from the \E8\x\E8 root lattice of the SUSY heterotic string, see section~\ref{sec:orbis}. In the model files, the name that appears in the lattice part is \E8 \x \E8, indicating this origin.

  \item The shift vectors. They are the 16D shift vectors $V_0, V_1$ and $V_2$ for an orbifold with point group $\Z{2W}\x\Z{M}\x\Z{N}$, or $V_0, V_1$ for $\Z{2W}\x\Z{M}$. The Witten shift vector is always $V_0 = (1,0^7,1,0^7)$ and corresponds to \Z{2W}, while $V_1$ and $V_2$ correspond $\Z{M}\x\Z{N}$. The shift vectors $V_0, V_1$ and $V_2$ represent the gauge embedding of the twist vectors $v_0, v_1$ and $v_2$.

  \item The Wilson lines. They are six 16D vectors denoted as $W_1,\ldots,W_6$, and they represent the gauge embedding of the 6D torus lattice vectors $e_\alpha$, $\alpha = 1,2,\ldots,6$. For example, for the \Z3 orbifold (i.e.\ \Z{2W}\x\Z3) with standard embedding and without Wilson lines, the model file is named \code{modelZ3\_1\_1.txt}, and the Wilson lines are null vectors. However, when an orbifold model with Wilson lines is created and saved to a model file, non-trivial Wilson lines appear in this part of the model files, see~\ref{subs:maindir}.
\end{itemize}

\subsubsection{A note for the names of the geometry and model files}
\label{subs:notenames}

{To study orbifold models with point group \Z{K}\x\Z{M}\x\Z{N}, where \Z{K}=\Z{2W} is the freely acting group, the corresponding geometry and model files that we provide are automatically} named \code{Geometry\_ZMxZN\_i\_j.txt} and \code{modelZMxZN\_i\_j.txt}, respectively. We follow the traditional convention of using the {label of the} point group {of the} compactification, i.e.\ \Z{M}\x\Z{N}, {as the core of the name of} these files. {These labels follow the known classification of 138 consistent Abelian orbifold geometries for orbifold compactifications of the heterotic string~\cite{Fischer:2012qj}} that lead to a vanishing cosmological constant~\cite{GrootNibbelink:2017luf}: 119 with \Z{M}\x\Z{N} point group, and 19 with \Z{M} point group. The standard notation of the orbifold geometries (space groups) is $\Z{M}\x\Z{N}\,(i,j)$, where $i$ and $j$ are positive integer numbers denoting the type of the 6D torus lattice and the presence of roto-translations, respectively. Specifically, $i = 1$ indicates a factorizable torus lattice, and $i > 1$ a non-factorizable lattice. {Also, only $j = 1$ indicates the absence of roto-translations.} For example, the \Z{3}\x\Z{3} (2,3) orbifold geometry indicates a non-factorizable lattice and the presence of roto-translations. {Note that for orbifolds with point group \Z{M}, none of the 19 orbifold geometries admits roto-translations and hence $j=1$ for all of them. So, in this case}, the corresponding names for the geometry and model files are \code{Geometry\_ZM\_i\_1.txt} and \code{modelZM\_i\_1.txt}, respectively. {In the cases where two different point groups have identical orders, we add the labels I and II to the corresponding space groups and, thus, the respective model files are named as \code{modelZMxZN-I\_i\_j.txt} and \code{modelZMxZN-II\_i\_j.txt}.}

%%%%%%%%%%%%%%%%%%% NEW SECTION %%%%%%%%%%%%%%%%%%%
\section{Creating models from scratch}
\label{sec:fromzero}

{In this section we walk the reader through the creation and analysis of two sample orbifold models}. The first one is a model based on the \Z3 (1,1) orbifold geometry, where we consider the standard embedding without Wilson lines, see also~\cite[\S Complementary notes, Table 3]{website:2025} for a sample input and output. The second one {is based on} the \Z3\x\Z3 (1,1) orbifold geometry {and the chosen set of shift vectors and Wilson lines lead to a} SM-like model (see section~\ref{subs:smlike}), {i.e.\ it exhibits potentially realistic features, which shall be explored in detailed elsewhere}. The geometry files of these orbifolds, as presented in section~\ref{subs:geomf} and {made available} in the \code{Geometry} folder, are named \code{Geometry\_Z3\_1\_1.txt} and \code{Geometry\_Z3xZ3\_1\_1.txt}, respectively. In {both of these models}, the six Wilson lines must fulfill the identifications $W_1=W_2$, $W_3=W_4$ and $W_5=W_6$.

\subsection{\Z3 orbifold model with the standard embedding and no Wilson lines}
\label{subs:z3model}

%standard embedding
Let us start by showing how to create a \Z{3} orbifold model with standard embedding and no Wilson lines {with the help of the \orbifolder{}. We open a terminal window and enter the folder where the \orbifolder{} is installed. The program is started} by typing
\begin{lstlisting}[style=consola,numbers=none]
$ ./nonSUSYorbifolder
\end{lstlisting}
Then, in the main directory of the program, we enter
\begin{lstlisting}[style=consola,numbers=none]
> create orbifold(Z3) with point group(3,1)
\end{lstlisting}
where \code{Z3} is the chosen name or label for this orbifold model. The {numbers \code{(3,1)} indicate that the chosen} point group corresponds to $\Z{M}\x\Z{N} = \Z{3}$, i.e. $M=3$ and $N=1$, where $N=1$ specifies the absence of a \Z{N} group. The command \code{create orbifold(Z3) with point group(3)} can be used for the same purpose. Next, enter the newly created orbifold directory\footnote{The orbifold directory is named after the chosen orbifold label.} named \code{Z3} by typing
\begin{lstlisting}[style=consola,numbers=none]
> cd Z3
\end{lstlisting}
{The \orbifolder{} displays four steps, which demand additional input,} to fully define the model. Entering the first {command}
\begin{lstlisting}[style=consola,numbers=none]
/Z3/model> print available space groups
\end{lstlisting}
leads to the orbifold geometries that are compatible with the chosen point group. In the \Z3 case, there is only one space group and the \orbifolder{} takes it automatically, {skipping the second step. At this stage}, the 6D torus lattice, the twist vector and the relations among the six Wilson lines are set. The remaining two steps are to input the shift vector and Wilson lines. We are interested in the standard embedding in this example{, which requires no Wilson lines. With} this purpose, we just use
\begin{lstlisting}[style=consola,numbers=none]
/Z3/model> set shift standard embedding
\end{lstlisting}
Then, the conditions of modular invariance are verified and the massless spectrum of this orbifold model is computed. {We could in principle include non-zero Wilson lines, but we omit the last step as we want to focus on pure standard embedding here.} Let us explore some properties of the created model. For example, to inspect some details of the 4D massless spectrum, we go to the \code{spectrum} directory. As the current directory is \code{model}, we go back to the orbifold directory,
\begin{lstlisting}[style=consola,numbers=none]
/Z3/model> cd ..
\end{lstlisting}
and enter
\begin{lstlisting}[style=consola,numbers=none]
/Z3> cd spectrum
\end{lstlisting}
to access the \code{spectrum} directory. The user can also use the shortcut \code{s} in the \code{model} directory to go directly to the \code{spectrum} directory (\code{/Z3/model/> s}). To see a list of the commands that are defined {to study the spectrum of the model}, type
\begin{lstlisting}[style=consola,numbers=none]
/Z3/spectrum> dir
\end{lstlisting}
The command \code{dir} {(or \code{help} with no modifiers) provides the required information; see~\ref{subs:spectdir} for more details on the commands of this directory}. The command
\begin{lstlisting}[style=consola,numbers=none]
/Z3/spectrum> help print summary
\end{lstlisting}
shows all the options that can be used with the command \code{print summary}. {For instance}, to print the 4D massless spectrum with field labels, we use the command
\begin{lstlisting}[style=consola,numbers=none]
/Z3/spectrum> print summary with labels
\end{lstlisting}
The information displayed consists of the 4D gauge group,\footnote{Usually, the 4D gauge group of heterotic orbifold models contains some \U1 factors. One of them can be (pseudo-)anomalous; the anomaly is canceled by the Green-Schwarz mechanism~\cite{Green:1984sg}.} the scalar and fermion fields characterized by their representations under the 4D gauge group, and the field labels in the current {configuration of field labels (and \U1 generator basis)}, which we call {\it vev-configuration}. In the {\Z3 orbifold with standard embedding}, the 4D gauge group is $\SO{10} \x \SU{3} \x \SO{16} \x \U1$ and the current vev-configuration is automatically labeled \code{TestConfig1}. The field labels are \code{s\_i} and \code{f\_j} for scalar and fermion fields, respectively, and the indices $i$ and $j$ count and uniquely identify these fields. The spectrum omitting all labels and \U1 charges can be printed by using the command
\begin{lstlisting}[style=consola,numbers=none]
/Z3/spectrum> print summary no U1s
\end{lstlisting}
{In some cases it is convenient to use the command}
\begin{lstlisting}[style=consola,numbers=none]
/Z3/spectrum> print summary of sectors
\end{lstlisting}
which shows the massless spectrum, split in the untwisted and the various twisted sectors.
To learn all details for a specific field, for example the fermion field labeled \code{f\_23}, type
\begin{lstlisting}[style=consola,numbers=none]
/Z3/spectrum> print(f_23)
\end{lstlisting}
This command shows the sector $(k,m,n)$, the fixed point by providing the numbers $n_\alpha$ (indicating $n_\alpha e_\alpha, \alpha = 1,2,\ldots,6$), the representations under the 4D gauge group, the left-moving momenta, the right-moving momenta, and the number of oscillators. To display this information for all fields instead, type
\begin{lstlisting}[style=consola,numbers=none]
/Z3/spectrum> print(*)
\end{lstlisting}
{It is possible to obtain the spectrum in \LaTeX{} format including all fields} by entering
\begin{lstlisting}[style=consola,numbers=none]
/Z3_1_1/spectrum> tex table(*) print labels(1) to file(textable.tex)
\end{lstlisting}
The resulting latex code (saved in the file \code{textable.tex} for our example) can be compiled to get a table that shows the scalar and fermion fields, the untwisted and twisted sectors, the representations and charges of the fields under the 4D gauge group, and the field labels.

{It is sometimes necessary or convenient to assign personalized labels to all fields. One can do so} in the \code{vev-config/labels} directory. We can use the shortcut
\begin{lstlisting}[style=consola,numbers=none]
/Z3/spectrum> l
\end{lstlisting}
to go directly to the \code{vev-config/labels} directory. A list of the commands in this directory can be {obtained} by using the command \code{dir} (or \code{help}). See~\ref{subs:labeldir} for more information on these commands. In particular, to change the current field labels use the command
\begin{lstlisting}[style=consola,numbers=none]
/Z3/vev-config/labels> create labels
\end{lstlisting}
{Then, each scalar and fermion in the spectrum is listed one by one, prompting the user to provide a new label for each field.} Now, use the command
\begin{lstlisting}[style=consola,numbers=none]
/Z3/vev-config/labels> print labels
\end{lstlisting}
to see that the new labels are already assigned for the fields in the spectrum. The message \code{Using label \#2 of the fields} also appears at the top. The labels stored as \code{\#1} correspond to the original, automatically generated field labels; i.e.\ \code{s\_i} and \code{f\_j} for scalar and fermion fields, respectively. To use the standard labels again, use the command
\begin{lstlisting}[style=consola,numbers=none]
/Z3/vev-config/labels> use label(1)
\end{lstlisting}
and execute again the command \code{print labels} to confirm that all fields have the original labels \code{s\_i} and \code{f\_j}.

Now, in the \code{model} directory we can explore some properties of the orbifold geometry. {We reach that directory by using} the shortcut
\begin{lstlisting}[style=consola,numbers=none]
/Z3/vev-config/labels> m
\end{lstlisting}
{Again, we type \code{dir} (or \code{help}) to know the commands in this directory; see~\ref{subs:modeldir} for details on those commands. We enter \code{help print} to see a list of all properties available from the command} \code{print}. For example, to obtain the twist and shift vectors of the model, we use the commands
\begin{lstlisting}[style=consola,numbers=none]
/Z3/model> print twist
\end{lstlisting}
and
\begin{lstlisting}[style=consola,numbers=none]
/Z3/model> print shift
\end{lstlisting}
Note that the first three components of the shift vector correspond to the first non-zero components of the twist vector, which is the characteristic of the standard embedding. To retrieve the point group of the orbifold model, we use the command
\begin{lstlisting}[style=consola,numbers=none]
/Z3/model> print point group
\end{lstlisting}
Other {possible directives based on the command \code{print} include} \code{print orbifold label} and \code{print \#SUSY}. The first one displays the given label of the model; which reads \code{Orbifold "Z3"} in this case. The second one displays \code{N = 0 SUSY in 4d}, indicating that our model is a non-SUSY model in 4D.

To quit the \orbifolder{}, we type \code{exit} and then \code{yes}, to confirm.

\subsection{A \Z3\x\Z3 orbifold model with Wilson lines}
\label{subs:z3z3model}

{Let us now study a phenomenologically viable model based on the \Z3\x\Z3 point group.} There are 15 orbifold geometries with {this point group}~\cite{Fischer:2012qj}. We {select the simplest orbifold geometry of this type, \Z3\x\Z3 (1,1). The corresponding geometry file is \code{Geometry\_Z3xZ3\_1\_1.txt}. As before, in a terminal window, in the installation folder, we start the \orbifolder{} by typing}
\begin{lstlisting}[style=consola,numbers=none]
$ ./nonSUSYorbifolder
\end{lstlisting}
A \Z3\x\Z3 orbifold model labeled \code{modelz3xz3} is initiated by the command
\begin{lstlisting}[style=consola,numbers=none]
> create orbifold(modelz3xz3) with point group(3,3)
\end{lstlisting}
Next, we enter the orbifold model directory with
\begin{lstlisting}[style=consola,numbers=none]
> cd modelz3xz3
\end{lstlisting}
{The \orbifolder{} displays the four steps that are required to fully define the model. First, we find the available space groups associated with the chosen point group}
\begin{lstlisting}[style=consola,numbers=none]
/modelz3xz3/model> print available space groups
\end{lstlisting}
{In our \Z3\x\Z3 case, a list of 15 geometries is shown. Let us choose the first space group corresponding to the \Z3\x\Z3 (1,1) geometry,}
\begin{lstlisting}[style=consola,numbers=none]
/modelz3xz3/model> use space group(1)
\end{lstlisting}
This command {sets the 6D torus lattice, the twist vectors and the constraints on the Wilson lines associated with the chosen geometry. In the next two steps, we must} set explicitly the shift vectors and the Wilson lines. For the shift vectors there are two options: set the shifts in the standard embedding or set distinct shifts vectors. In this example, {we set non-standard-embedding shift vectors and define non-trivial Wilson lines. To set our chosen shift vectors,} we use the command\footnote{Note that this instruction is equivalent to (see~\ref{subsub:vect} for the various {\orbifolder{}} formats for vectors) \\\code{\footnotesize{set shift V(1) = (-1/6,-1/6,-1/6,-1/6,-1/6,-1/6,1/2,1/2,-11/6,-1/6,-1/6,1/6,1/6,1/6,1/6,5/6)}}}
\begin{lstlisting}[style=consola,numbers=none]
/modelz3xz3/model> set shift V(1)=(-1/6^6, 1/2^2, -11/6, -1/6^2, 1/6^4, 5/6)
\end{lstlisting}
and
\begin{lstlisting}[style=consola,numbers=none]
/modelz3xz3/model> set shift V(2)=(1/6^5,5/6,-11/6,1/6,1/2,-5/6,1/2,-1/6,1/6^2,13/6,-7/6)
\end{lstlisting}
{These vectors have been selected to be consistent. The \orbifolder{} verifies the consistency conditions of the shift vectors and, if fulfilled, computes the resulting 4D massless spectrum. Now, to define the Wilson lines, we must recall the identification conditions for the chosen geometry, which in our case are $W_1 = W_2$, $W_3 = W_4$ and $W_5 = W_6$. Hence, in an orbifold model with \Z3\x\Z3 (1,1) geometry, we must consider that there are only three independent Wilson lines, say $W_1$, $W_3$ and $W_5$. Let us take $W_1=W_5=0$ (which is the default) and set a non-vanishing $W_3$ by typing}
\begin{lstlisting}[style=consola,numbers=none]
/modelz3xz3/model>
set WL W(3) = (-11/6,-7/6,-1/6,-1/6,1/6,-7/6,5/6,5/6,11/6,7/6,-3/2,1/6,1/6,5/6,13/6,-1/6)
\end{lstlisting}
After executing the last command, the \orbifolder{} confirms the identified Wilson lines, verifies that modular invariance is still satisfied and computes the new 4D massless spectrum.

Now the creation of the orbifold model is complete and one can explore its properties by using the commands in the different directories. For example, to know if this model allows for vacua compatible with the {SM, or Pati-Salam (PS) or \SU5 GUTs, we can go to the \code{vev-config} directory and use the command \code{analyze config}, see~\ref{subs:vdir} for details of the commands in the \code{vev-config} directory.} Since the current directory is \code{model}, we use the shortcut $\code{v}$ to go directly to the \code{vev-config} directory, i.e.
\begin{lstlisting}[style=consola,numbers=none]
/modelz3xz3/model> v
\end{lstlisting}
and then write the command
\begin{lstlisting}[style=consola,numbers=none]
/modelz3xz3/vev-config> analyze config
\end{lstlisting}
{The \orbifolder{} lets us know} that this orbifold model allows for SM vacua. It also shows the massless spectrum and assigns automatically proper labels for the scalar and fermion fields; for example, \code{l\_1}, \code{l\_2} and \code{l\_3} {denote the three SM} lepton doublets. The spectrum has the {SM gauge group, ${\mathcal{G}}_\text{SM}= \SU{3}_c\x\SU{2}_L\x\U1_Y$, and SM matter spectrum, enlarged with some exotic fields}. They are presented according to their SM gauge representations and charges.

To quit the \orbifolder{} type \code{exit} and then \code{yes}.

\section{Creating random models}
\label{sec:randommodels}
%SM-like model

In this section, we demonstrate how to use the \orbifolder{} to randomly create classes of SM-like models. Our search is based on the \Z2\x\Z4 (1,6) orbifold geometry,\footnote{For higher orders of the point group, such as \Z{12} or \Z6\x\Z6, it is required to enlarge the parameter space (or lattice) used to create admissible shifts and Wilson lines. This is done by editing the function \texttt{Initiate} in the source-program file \texttt{crandommodel.cpp} (located in the folder \texttt{src/orbifolder}). After code line 217, one must choose a wider region where the parameters $a_i$ run or introduce a deeper \texttt{for} series. A (commented) example for this is included starting at code line 247.} which is known to produce many promising models~\cite{Perez-Martinez:2021zjj}. We start the \orbifolder{} {by executing the program in a terminal}
\begin{lstlisting}[style=consola,numbers=none]
$ ./nonSUSYorbifolder
\end{lstlisting}
{To load the orbifold model defined in the available file \code{modelZ2xZ4\_1\_6.txt}, we type}
\begin{lstlisting}[style=consola,numbers=none]
> load orbifolds(Models/ZNxZM_models/modelZ2xZ4_1_6.txt)
\end{lstlisting}
{The crucial instruction to initiate the random creation of SM-like models from this orbifold reads}\footnote{The complete instruction must be written in one line, i.e.\ as\\
  \scriptsize{create random orbifold from(Z3\_1\_1) if(inequivalent) \#models(8) use(0,0,1,0,0,0,1,0) save to(models.txt) print info load when done}}
\begin{lstlisting}[style=consola,numbers=none]
> create random orbifold from(Z2xZ4_1_6) if(inequivalent SM) #models(3)
            use(0,0,0,0,0,0,0,0) save to(models.txt) print info load when done
\end{lstlisting}
{As indicated by the modifier \code{\#models(3)} and \code{if(inequivalent SM)},} this instruction creates three inequivalent\footnote{Two orbifold models are equivalent {if the non-Abelian quantum numbers and the number of singlets in their spectra coincide.}} SM-like models with \Z2\x\Z4 (1,6) orbifold geometry. The {modifier \code{use(0,0,0,0,0,0,0,0)} indicates that none of the original shift vectors and Wilson lines is used in the newly created models, i.e.\ the} shift vectors $V_1$ and $V_2$ and the six Wilson lines $W_\alpha$ are created randomly. {After creating these vectors,} the modular invariance conditions~\eqref{eq:modInv} are checked. {If they lead to an admissible model, the created data is stored. When all three models are correctly created,} the models are saved to a file named \code{models.txt}, a short summary of the spectrum is printed, and, finally, the models are loaded into three orbifold directories named \code{Model\_SM1}, \code{Model\_SM2} and \code{Model\_SM3}. For more information about the parameters used with the command \code{create random orbifold from(OrbifoldLabel)}, see~\ref{subs:maindir}. To enter {e.g.\ the orbifold directory} \code{Model\_SM1}, we write
\begin{lstlisting}[style=consola,numbers=none]
> cd Model_SM1
\end{lstlisting}
One can now explore the properties of the orbifold model by choosing any of the directories (\code{model, gauge group, spectrum, vev-config} and \code{vev-config/labels}) and the commands defined therein. For example, details of the orbifold geometry such as the twist vectors, shift vectors and Wilson lines that define the model, {can be accessed by entering} the \code{model} directory
%% model directory SM1 TC1
\begin{lstlisting}[style=consola,numbers=none]
/Model_SM1> cd model
\end{lstlisting}
and using
\begin{lstlisting}[style=consola,numbers=none]
/Model_SM1/model> print twists
\end{lstlisting}
to display the twist vectors,
\begin{lstlisting}[style=consola,numbers=none]
/Model_SM1/model> print shifts
\end{lstlisting}
{to read the randomly created shift vectors, and}
\begin{lstlisting}[style=consola,numbers=none]
/Model_SM1/model> print Wilson lines
\end{lstlisting}
for the Wilson lines, including the identification properties of the Wilson lines and their orders. The orbifold labels are obtained via the command
\begin{lstlisting}[style=consola,numbers=none]
/Model_SM1/model> print orbifold label
\end{lstlisting}
which shows the label \code{Model\_SM1} in this case. The point group of the orbifold model can be displayed by using the command
\begin{lstlisting}[style=consola,numbers=none]
/Model_SM1/model> print point group
\end{lstlisting}
{that yields} \code{Point group is Z\_2 x Z\_4}. The directive
\begin{lstlisting}[style=consola,numbers=none]
/Model_SM1/model> print #SUSY
\end{lstlisting}
{produces \code{N = 0 SUSY in 4d}, as expected.}

%%gauge group
In the \code{gauge group} directory the user can know different details of the 4D gauge group. We use the shortcut \code{gg} to go directly to the \code{gauge group} directory
\begin{lstlisting}[style=consola,numbers=none]
/Model_SM1/model> gg
\end{lstlisting}
We can see the 4D gauge group of the orbifold model by typing
\begin{lstlisting}[style=consola,numbers=none]
/Model_SM1/gauge group> print gauge group
\end{lstlisting}
Then, the 4D gauge group in the current vev-configuration, \code{TestConfig1} at this step, is shown. A choice of simple roots for the 4D gauge group can be seen by using the command
\begin{lstlisting}[style=consola,numbers=none]
/Model_SM1/gauge group> print simple roots
\end{lstlisting}
{Other information of interest can be retrieved by using various additional commands available in the directory, which can be accessed by using the commands \code{help} and \code{help print}. This includes the beta-function coefficients, the gauge and gravitational anomalies, and the details on the basis of the gauge \U1s.

Details on the matter spectrum of an orbifold model can be obtained in the \code{spectrum} directory. To get there, we use the shortcut}
\begin{lstlisting}[style=consola,numbers=none]
/Model_SM1/gauge group> s
\end{lstlisting}
followed by the instruction
\begin{lstlisting}[style=consola,numbers=none]
/Model_SM1/spectrum> print summary with labels
\end{lstlisting}
{where the optional modifier \code{with labels} has been invoked to visualize the labels given automatically to scalar and fermion fields.
The massless spectrum is displayed in terms of the field gauge quantum numbers under the full 4D gauge group $G_{4D}$. The standard labels accompanying the gauge representations are \code{s\_i} for scalars and \code{f\_j} for fermions, where the indices \code{i} and \code{j} count the number of them. Other available commands in this directory can be viewed by executing \code{help} and \code{help print summary}.

%vev-config
These labels \code{s\_i} and \code{f\_j} are stored in the automatic vev-configuration \code{TestConfig1}. Since the model is constructed to exhibit SM-like properties, this field labeling is not ideal. We thus seek to identify the fields corresponding to the quarks and leptons of the SM more accurately. To systematically label the SM fields, it is necessary to access the \code{vev-config} directory by e.g.\ using the shortcut}
\begin{lstlisting}[style=consola,numbers=none]
/Model_SM1/gauge group> v
\end{lstlisting}
Once there, we execute
\begin{lstlisting}[style=consola,numbers=none]
/Model_SM1/vev-config> analyze config
\end{lstlisting}
{which lets the \orbifolder{} analyze the gauge group and spectrum of the model, identifying whether it has the properties of the SM, or a PS or \SU5 GUT, including three generation of fermions and vector-like exotics. In the current model, it identifies a SM-like configuration, which is automatically labeled as \code{SMConfig1}.}
It also shows the massless spectrum in this vev-configuration and assigns automatically appropriate labels for the scalar and fermion fields, for example \code{h\_1,h\_2,\ldots} for (scalar) Higgs doublets and \code{l\_1,l\_2,l\_3} for the three generations of lepton doublets. The scalar and fermion fields are presented according to their representations and hypercharge under the SM gauge group ${\mathcal{G}}_\text{SM}= \SU{3}_c\x\SU{2}_L\x\U1_Y$, which {constitutes the observable sector in this vev-configuration. All other gauge-group factors are considered ``hidden'' and the associated non-Abelian quantum numbers are regarded as multiplicities of the presented fields. The command \code{analyze config} admits the optional modifier \code{print SU(5) simple roots}, which additionally provides a subset of \SO{16} simple roots building an \SU5 in which ${\mathcal{G}}_\text{SM}$ is embedded, ensuring the compatibility of the hypercharge $Y$ with \SU5 unification at higher energies. (If required, in the \code{gauge group} directory all simple roots of non-Abelian gauge factors and \U1 generators of $G_{4D}$ can be retrieved by \code{print simple roots} and \code{print U1 generators}, respectively. The hypercharge is identified with the label \code{Y}.) To display the gauge group with the current selection of observable and hidden gauge sector, we enter the command}
\begin{lstlisting}[style=consola,numbers=none]
/Model_SformsM1/vev-config> print gauge group
\end{lstlisting}
The \orbifolder{} prints out the full 4D gauge group, where group factors in brackets belong to the hidden sector.
We shall shortly explore other tools in the \code{vev-config} directory, but let us now see how the changes in the vev-configuration are reflected in the \code{spectrum} directory.

After changing to the \code{spectrum} directory with e.g.
\begin{lstlisting}[style=consola,numbers=none]
/Model_SM1/vev-config> s
\end{lstlisting}
we execute the command
\begin{lstlisting}[style=consola,numbers=none]
/Model_SM1/spectrum> print summary with labels
\end{lstlisting}
The 4D massless spectrum is displayed in the vev-configuration \code{SMConfig1}. It is also possible to learn whether matter fields arise from the untwisted or twisted sectors, by using the command
\begin{lstlisting}[style=consola,numbers=none]
/Model_SM1/spectrum> print summary of sectors
\end{lstlisting}
{One might prefer to explore the spectrum arising in a particular sector \code{T(k,m,n)}. For example,}
\begin{lstlisting}[style=consola,numbers=none]
/Model_SM1/spectrum> print summary of sector T(1,0,2)
\end{lstlisting}
For the untwisted sector use the command \code{print summary of sector T(0,0,0)}. {The previous commands also admit the modifier \code{with labels} for better identification of SM fields.} Additional information for the fields can be obtained by using the command \code{print(fields)}. For instance, for {a lepton doublet} labeled as \code{l\_1}, enter
\begin{lstlisting}[style=consola,numbers=none]
/Model_SM1/spectrum> print(l_1)
\end{lstlisting}
{Among other features, this provides the details of the localization via the constructing element of the original string, the ${\mathcal{G}}_\text{SM}$ representations (as well as the representations under the full $G_{4D}$ gauge group), and the left- and the right-moving momenta associated to the field \code{l\_1}.}

Let us now explore more advanced operations to perform in the \code{vev-config} directory. We go back to the \code{vev-config} directory,
\begin{lstlisting}[style=consola,numbers=none]
/Model_SM1/spectrum> v
\end{lstlisting}
{and display all available vev-configurations with the command}
\begin{lstlisting}[style=consola,numbers=none]
/Model_SM1/vev-config> print configs
\end{lstlisting}
The current configuration is marked by an arrow, which in this case is \code{SMConfig1}. {Other configurations include \code{StandardConfig1} and \code{TestConfig1}. We can choose our working configuration by using e.g.}
\begin{lstlisting}[style=consola,numbers=none]
/Model_SM1/vev-config> use config(TestConfig1)
\end{lstlisting}
which lets the \orbifolder{} focus on the \code{TestConfig1} vev-configuration. It shows the message \code{Now using vev-configuration "TestConfig1"}. At this step, the full 4D gauge group is regarded as observable gauge sector, as by default. To confirm this, we execute
\begin{lstlisting}[style=consola,numbers=none]
/Model_SM1/vev-config> print gauge group
\end{lstlisting}
where the 4D gauge group is shown without any group factors in brackets. The \code{TestConfig1} contains both the default 4D gauge group and labels for matter fields. If we want to restore the identified SM field labels, we can reset the \code{SMConfig1} configuration via
\begin{lstlisting}[style=consola,numbers=none]
/Model_SM1/vev-config> use config(SMConfig1)
\end{lstlisting}
This assigns \code{SMConfig1} as the current vev-configuration for the orbifold model. At this step, the observable sector is {again} ${\mathcal{G}}_\text{SM}= \SU{3}_c\x\SU{2}_L\x\U1_Y$. Let us now show how to change the observable gauge sector. First, it is recommended to use the command
\begin{lstlisting}[style=consola,numbers=none]
/Model_SM1/vev-config> print gauge group
\end{lstlisting}
to identify the position number of the gauge group factors in $G_{4D}$ and to know the groups that form the hidden and the observable gauge sectors. Of course, the SM gauge group is the current observable sector. To change the observable gauge sector, we use the command \code{select observable sector:\,[parameters]}, see~\ref{subs:vdir} for details. For example,
\begin{lstlisting}[style=consola,numbers=none]
/Model_SM1/vev-config> select observable sector: gauge group(2,3) U1s(2,3)
\end{lstlisting}
where the numbers \code{(2,3)} after \code{gauge group} indicate the position of the non-Abelian gauge group factors in $G_{4D}$. Similarly, the numbers \code{(2,3)} after \code{U1s} indicate the position of the \code{U1s}. To illustrate the previous example, suppose that, initially, $G_{4D}$ in \code{SMConfig1} reads
$$[\SU{5}] \x \SU{3}_c \x \SU{2}_L \x [\SU{2}] \x [\U1_1] \x \U1_{2,Y} \x [\U1_3] \x \ldots \x [\U1_8]\,,$$
where factors in brackets belong to the hidden sector. Clearly, the observable sector is the SM group, where $\SU{3}_c$ and $\SU{2}_L$ have, respectively, the positions 2 and 3 in the non-Abelian part. The hypercharge $\U1_{2,Y}$ has the position 2 in the Abelian part. So, the command \code{select observable sector:\,gauge group(2,3) U1s(2,3)} assigns the group
$$\SU{3}_c\x\SU{2}_L\x\U1_Y \x \U1_3$$
as the new observable sector, which can be useful in {e.g.\ studies of models with extra $Z'$-like Abelian interactions}.
One can find more information about the \code{select} command and further examples in its manual pages, which are accessed via
\begin{lstlisting}[style=consola,numbers=none]
/Model_SM1/vev-config> man select
\end{lstlisting}
{Once the observable sector is changed, one can print the corresponding spectrum by using the command \code{print summary with labels} in the \code{spectrum} directory. The SM-labels are preserved even though the observable gauge sector is changed. One might prefer to have this configuration in a copy with a different name. For that purpose, we use the command}
\begin{lstlisting}[style=consola,numbers=none]
/Model_SM1/vev-config> create config(ZprimeConfig) from(SMConfig1)
\end{lstlisting}
which creates the new vev-configuration \code{ZprimeConfig} using the current choice of labels and observable gauge sector in the \code{SMConfig1} configuration, including the extra \U1 factor.

\subsection{Searches of promising models with the \orbifolder{}}

Following the techniques just explained, the \orbifolder{} has been used for performing random searches of SM-like models~\cite{Blaszczyk:2014qoa,Perez-Martinez:2021zjj}. For example, in ref.~\cite{Blaszczyk:2014qoa} a search of these models on selected orbifold geometries was realized and models with one Higgs doublet were found. In ref.~\cite{Perez-Martinez:2021zjj} all 138 Abelian orbifold geometries were considered and led to about 170,000 SM-like models, available at the website~\cite{website:2021}. By inspecting the massless spectrum of these models, exotic particles were classified and the best SM-like were identified. These searches led to a study were dark-matter candidates were identified in promising non-SUSY orbifold compactifications~\cite{Cervantes:2023wti}. These constructions can now be further explored by using the publicly available version of the \orbifolder{}.

%%%%%%%%%%%%%%%%%%%%%%%%%%%%%%%%%%%%%%%%%%%%%%%%%%%%%%%%%%%%%%%%%%%%%%%
\section{Creating and loading \SU5 GUT models}
\label{sec:loadmodels}
%SU5
In this section we {describe the steps to randomly create orbifold models with properties of \SU5 GUTs based on the \Z3 (1,1) geometry. We shall store them in a file, which will then be used to show the tool to load previously built models and study their properties.}

Let us start by creating and saving to a file three inequivalent \SU5 models. {First, in a terminal window within the installation folder, we start the \orbifolder{} via the command}
\begin{lstlisting}[style=consola,numbers=none]
$ ./nonSUSYorbifolder Models/ZN_models/modelZ3_1_1.txt
\end{lstlisting}
We can now create the models we need by typing\footnote{It is required to write the complete instruction in one line, i.e.\\
 \footnotesize{create random orbifold from(Z3\_1\_1) if(inequivalent SU5) \#models(3) use(0,0,0,0,0,0,0,0) save to(modelsSU5.txt) print info}}
\begin{lstlisting}[style=consola,numbers=none]
> create random orbifold from(Z3_1_1) if(inequivalent SU5) #models(3)
    use(0,0,0,0,0,0,0,0) save to(modelsSU5.txt) print info
\end{lstlisting}
The \orbifolder{} creates three inequivalent \SU5 models, where the shift vector and the Wilson lines are created randomly, the models are saved to a file named \code{modelsSU5.txt} and a brief summary of the spectrum is printed {thanks to the modifier \code{print info}}. Now, let us quit the \orbifolder{} by typing \code{exit} and \code{yes}.

{To load the previously constructed \SU5 GUT models from the file \code{modelsSU5.txt}, the software can be restarted using the command}
\begin{lstlisting}[style=consola,numbers=none]
$ ./nonSUSYorbifolder modelsSU5.txt
\end{lstlisting}
The three orbifold models are loaded and stored in the directories \code{Model\_SU5\_1, Model\_SU5\_2} and \code{Model\_SU5\_3}, as can be confirmed by using
\begin{lstlisting}[style=consola,numbers=none]
> dir
\end{lstlisting}
{Now, we enter e.g.\ the \code{Model\_SU5\_3} orbifold directory:}
\begin{lstlisting}[style=consola,numbers=none]
> cd Model_SU5_3
\end{lstlisting}
To know the shift vector and the Wilson lines, we go the \code{model} directory
\begin{lstlisting}[style=consola,numbers=none]
/Model_SU5_3> m
\end{lstlisting}
and use the commands
\begin{lstlisting}[style=consola,numbers=none]
/Model_SU5_3/model> print shift
\end{lstlisting}
and
\begin{lstlisting}[style=consola,numbers=none]
/Model_SU5_3/model> print Wilson lines
\end{lstlisting}
To see the 4D gauge group, {we access} the \code{gauge group} directory
\begin{lstlisting}[style=consola,numbers=none]
/Model_SU5_3/model> gg
\end{lstlisting}
and type
\begin{lstlisting}[style=consola,numbers=none]
/Model_SU5_3/gauge group> print gauge group
\end{lstlisting}
Then, the 4D gauge group is printed in the current vev-configuration \code{TestConfig1}. As none of the group factors appears in brackets, the observable gauge sector consists on the full 4D gauge group. The beta coefficients for the non-Abelian gauge group factors of the observable sector can be obtained with the command
\begin{lstlisting}[style=consola,numbers=none]
/Model_SU5_3/gauge group> print beta coefficients
\end{lstlisting}
To know the simple roots and \U1 generators type
\begin{lstlisting}[style=consola,numbers=none]
/Model_SU5_3/gauge group> print simple roots
\end{lstlisting}
and
\begin{lstlisting}[style=consola,numbers=none]
/Model_SU5_3/gauge group> print U1 generators
\end{lstlisting}
To see the 4D massless spectrum, we go to the \code{spectrum} directory
\begin{lstlisting}[style=consola,numbers=none]
/Model_SU5_3/gauge group> s
\end{lstlisting}
and type, for example,
\begin{lstlisting}[style=consola,numbers=none]
/Model_SU5_3/spectrum> print summary with labels
\end{lstlisting}
Then, the scalar and fermion fields are printed with their representations and charges under the 4D gauge group. The current vev-configuration is \code{TestConfig1}, where the default labels \code{s\_i} and \code{f\_j} for the scalars and fermions are used. To get the \SU5 vev-configuration of this model, we go to the \code{vev-config} directory
\begin{lstlisting}[style=consola,numbers=none]
/Model_SU5_3/spectrum> v
\end{lstlisting}
and type
\begin{lstlisting}[style=consola,numbers=none]
/Model_SU5_3/vev-config> analyze config
\end{lstlisting}
Thereby, the \orbifolder{} identifies the \SU5 vev-configuration, named \code{SU5Config1}, for this orbifold model. The spectrum is also printed, where the scalar and fermion fields appear with their representations under the \SU5 gauge group, which {now builds the observable gauge sector}. A new set of labels, compatible with \SU5 GUTs, are also assigned to the fields.

Following the techniques explained in previous sections, we can freely proceed to further explore and study from the various subdirectories the properties of the model(s) and/or produce more interesting models. For example, if we wanted to produce a large set of models instead of only three, we may replace the modifier \code{\#models(3)} by \code{\#models(all)}. Although we have explained the use of most of the instructions of the \orbifolder{}, it is advisable to use our glossary in~\ref{app:commands} as a reference tool to support all analyses conducted with our software.

%%%%%%%%%%%%%%%%%%% NEW SECTION %%%%%%%%%%%%%%%%%%%
\section{Conclusions and outlook}
\label{sec:conclusions}

Since the discovery of the Higgs particle by the LHC, which does not exhibit supersymmetric properties yet,
there has been a renewed interest in model building without SUSY. Compactifications of string theory offer
a potential ultraviolet completion of such models, which could also produce predictions based on its
top-down ingredients, such as extra dimensions, particles and forces. These properties have already
been successfully explored in the literature based on orbifold compactifications of the non-SUSY heterotic string,
displaying promising results. However, a proper search requires an automatized tool allowing an extensive
search and analysis of models capable of reproducing observations in particle physics and
cosmology, and of producing new phenomenological features to contrast with current and future probes.

In this work, we introduced the \orbifolder{}, a useful software designed to generate and analyze 4D non-SUSY, tachyon-free
string models that arise from the compactification of the non-SUSY heterotic string on Abelian toroidal orbifolds. To build
a non-SUSY heterotic orbifold, some input parameters are required, including the space group defining the orbifold,
its embedding (including shifts and Wilson lines), and the gauge degrees of freedom. These parameters are provided
to the \orbifolder{}, which can both accept them from a user and generate them automatically.

Using this data, the program produces the effective 4D gauge group, including details such as the simple roots,
beta-function coefficients, \U1 generators, and associated gauge anomalies. It also calculates the low-energy
massless spectrum of 4D bosons and fermions resulting from the compactification. Additionally, the \orbifolder{}
can automatically generate models that reproduce the gauge symmetry and spectrum of the SM, as well as extensions
with GUT-compatible gauge groups such as \SU5.

The import/export commands allow the user to study previously generated data and to save information for future
analysis. Further, the \orbifolder{} can prepare \LaTeX{} code useful for presenting the resulting data.

The \orbifolder{} is programmed in \texttt{C++} and runs at best on a Linux-based system. However, we
provide also a Docker Container, which can be used in different platforms, such as Windows and Mac.

There is still a number of aspects that one might need to further investigate in order for them to be incorporated
in the \orbifolder{}, such as the modular and non-modular symmetries that may serve as flavor symmetries~\cite{Lauer:1989ax,Lauer:1990tm,Kobayashi:2006wq,Olguin-Trejo:2018wpw,Baur:2019iai,Nilles:2020gvu,Baur:2021mtl,Baur:2024qzo,Funakoshi:2025lxs} and
the somewhat related string selection rules that determine the admissible interactions in the constructed models~\cite{Kobayashi:2011cw,Nilles:2013lda,Dong:2025pah}.
More demanding questions involve the possibility of unstable vacua, including the possible non-perturbative appearance
of tachyons in the theory~\cite{Abel:2015oxa,GrootNibbelink:2017luf,Acharya:2020hsc}, and the inclusion of
non-Abelian~\cite{Konopka:2012gy,Fischer:2013qza} and asymmetric orbifolds~\cite{Narain:1986qm,Aldazabal:2025zht}.
These aspects are beyond the scope of the present work and will be studied elsewhere.

\section*{Acknowledgments}

It is a pleasure to thank Stefan Groot-Nibbelink, Orestis Lukas, Michael Blaszczyk
as well as Esa\'u Cervantes and Omar P\'erez-Figueroa for our
enlightening discussions and fruitful collaborations that led to the current work.
This work is partly supported by UNAM-PAPIIT IN113223 and Marcos Moshinsky Foundation.

\appendix
%\clearpage
%\newpage

\section{Glossary of commands}
\label{app:commands}

In this appendix we provide brief explanations for all commands of the prompt . In~\ref{sub:gralc} we present some concepts and general commands. In~\ref{sub:direct} we list the commands defined in each directory of the prompt. In~\ref{sub:man} we introduce the command \code{man}, which offers broad information and many examples about the prompt commands.

\subsection{Concepts and general commands}
\label{sub:gralc}

In this section we present some concepts and general commands that are useful for fields and scripts. The scalar and fermion fields are tagged with proper labels according to the current vev-configuration, and these labels are used to access some field details (\ref{subsub:fieldl}). A convenient way to deal with several fields is by defining a set of fields (\ref{subsub:setsf}). For some commands dealing with fields is possible to print details only for fields that satisfy certain conditions (\ref{subsub:ifcond}). We also describe the concept of processes (\ref{subsub:proc}), the use of vectors (\ref{subsub:vect}), how to change the output to \LaTeX{} or to Mathematica style (\ref{subsub:outputml}), and the use of system commands and variables (\ref{subsub:scv}).

\subsubsection{Field labels}
\label{subsub:fieldl}

Fields of the 4D orbifold models are tagged with labels according to the current vev-configuration. For example, an orbifold model in the vev-configuration \code{TestConfig1} or \code{StandardConfig1} the labels for the scalar and fermion fields are denoted as \code{s\_i} and \code{f\_j}, where \code{i} and \code{j} counts the total number of scalars and fermion fields, respectively. For models that allow SM vacua the vev-configuration is named \code{SMConfig1}. In this case, proper labels are assigned to the fields. For instance, \code{l\_1, l\_2} and \code{l\_3}, for the three generations of lepton doublets.

Several commands that print details for fields in the spectrum, like \code{print(fields)}, allow to access a field, a set of fields or all fields. For example, for an orbifold model in the vev-configuration \code{TestConfig1}, one can access the scalar field \code{s\_1} just by using \code{s\_1} (i.e. \code{print(s\_1)}), or \code{s\_1 s\_8} to access \code{s\_1} and \code{s\_8} scalar fields (i.e \code{print(s\_1 s\_8)}). To access all scalars except \code{s\_5} use \code{s-s\_5} (i.e. \code{print(s-s\_5)}). Similarly, to access all fermions except \code{f\_10} use \code{f-f\_10}. To access all fields use \code{*} (i.e. \code{print(*)}). Then, to access all scalars use \code{s}, or \code{*-f}. Similarly, to access all fermions use \code{f} or \code{*-s} (i.e. \code{print(f)} or \code{print(*-s)}). The explanation of the command \code{print(fields)} is given in~\ref{subs:spectdir}, where we present the commands of the \code{spectrum} directory.

The field labels are stored in the currently used vev-configuration of an orbifold model. The labels can be changed, except for models in the \code{StandardConfig1}. To create new field labels use the command \code{create labels} in the \code{vev-config/labels} directory. For more details, see~\ref{subs:labeldir}.

\subsubsection{Sets of fields}
\label{subsub:setsf}

It is possible to define sets of fields and use them on the same footing as fields. To work with sets of fields the following commands can be used.

\paragraph{\code{create set(SetLabel)}} This command creates an empty set named \code{SetLabel}.

\paragraph{\code{delete set(SetLabel)}} This command deletes the set named \code{SetLabel}.

\paragraph{\code{delete sets}} This command deletes all sets created.

\paragraph{\code{insert(fields) into set(SetLabel)}} This command inserts \code{fields} into a created set named \code{SetLabel}. It can be used with the parameter\\
\code{if(conditions)}\\
where only the fields that satisfy the condition are inserted in the set called \code{SetLabel}. Details for the parameter \code{if(conditions)} are presented in~\ref{subsub:ifcond}.

\paragraph{\code{remove(fields) from set(SetLabel)}} This command removes fields from the set named \code{SetLabel}. It can also be used with the parameter\\
\code{if(conditions)}\\
where only the fields that satisfy the condition are removed from the set called \code{SetLabel}.

\paragraph{\code{print sets}} This command shows all created sets. It can be used with the parameter\\
\code{if not empty}\\
In this case only the not empty sets are printed.

\paragraph{\code{print set(SetLabel)}} This command prints the set called \code{SetLabel}.

\paragraph{\code{\#fields in set(SetLabel)}} This command counts the number of fields in the set called \code{SetLabel}.

The following example shows the use of some of the previous commands in the \code{spectrum} directory. We use the \Z3 orbifold. In the directory where the program is installed write
\begin{lstlisting}[style=consola,numbers=none]
$ ./nonSUSYorbifolder modelZ3_1.txt
\end{lstlisting}
\begin{lstlisting}[style=consola,numbers=none]
> cd Z3_1_1
\end{lstlisting}
\begin{lstlisting}[style=consola,numbers=none]
/Z3_1_1> cd spectrum
\end{lstlisting}
Then, create a set named \code{test1}
\begin{lstlisting}[style=consola,numbers=none]
/Z3_1_1/spectrum> create set(test1)
\end{lstlisting}
then, to insert six fields into the set \code{test1} use the command
\begin{lstlisting}[style=consola,numbers=none]
/Z3_1_1/spectrum> insert(f_4 f_16 s_7 s_1 f_35 s_16) into set(test1)
\end{lstlisting}
to print the content of this set type
\begin{lstlisting}[style=consola,numbers=none]
/Z3_1_1/spectrum> print set(test1)
\end{lstlisting}
to display the number of fields in this set write
\begin{lstlisting}[style=consola,numbers=none]
/Z3_1_1/spectrum> #fields in set(test1)
\end{lstlisting}
to remove the fields with a non-zero number of left oscillators use
\begin{lstlisting}[style=consola,numbers=none]
/Z3_1_1/spectrum> remove(*) from set(test1) if(#osci. != 0)
\end{lstlisting}
this removes the field \code{s\_16}, as can be seen with the command
\begin{lstlisting}[style=consola,numbers=none]
/Z3_1_1/spectrum> print set(test1)
\end{lstlisting}
finally, to delete this set of fields use the command
\begin{lstlisting}[style=consola,numbers=none]
/Z3_1_1/spectrum> delete set(test1)
\end{lstlisting}

Additional details and examples of \code{sets} can be seen by typing \code{help sets} or \code{man sets} in the \code{spectrum} directory.

\subsubsection{Conditional \code{if}}
\label{subsub:ifcond}

Commands dealing with fields select only those fields that satisfy the condition. In general, the condition consists of three parts: the variable, the comparison operator, and the value. For example,
\begin{itemize}
  \item \code{if(length == 2/3)}. This selects the fields where the length-square of the shifted left-moving momentum is equal to \code{2/3}. Here the variable is \code{length}, the comparison operator is \code{==}, and the value is \code{2/3}.
  \item \code{if(Q\_1 == 12)}. This chooses the fields with the first \U1 charge equal to 12. Here the variable is \code{Q\_1}, the comparison operator is \code{==}, and the value is \code{12}.
  \item \code{if(\#osci.\,!= 0)}. This selects the fields where the number of oscillators acting on the left mover is non zero. In this case the variable is \code{\#osci.}, the comparison operator is \code{!=}, and the value is \code{0}.
  \item \code{if(q\_sh\_2 == -1/6)}. This chooses the fields where the second component of the shifted right-moving momentum is equal to \code{-1/6}. In this example the variable is \code{q\_sh\_2}, the comparison operator is \code{==}, and the value is \code{-1/6}.
\end{itemize}

As a brief example, consider the \Z3 orbifold. In the directory where the program is installed write
\begin{lstlisting}[style=consola,numbers=none]
$ ./nonSUSYorbifolder modelZ3_1.txt
\end{lstlisting}
\begin{lstlisting}[style=consola,numbers=none]
> cd Z3_1_1
\end{lstlisting}
\begin{lstlisting}[style=consola,numbers=none]
/Z3_1_1> cd spectrum
\end{lstlisting}
Then, use the command
\begin{lstlisting}[style=consola,numbers=none]
/Z3_1_1/spectrum> print(*) if(Q_1 == 24)
\end{lstlisting}
In this case, the information of the command \code{print(*)} (see~\ref{subs:spectdir}) is displayed but only for the fields that have the \U1 charge equal to 24. The 4D gauge group of this model is $\SO{10} \x \SU{3} \x \SO{16} \x \U1$. Commands like \code{print(fields)}, \code{tex table(fields)} and \code{print list of charges(fields)} in the \code{spectrum} directory can be used with the conditional \code{if}. More examples and details can be seen by typing \code{help conditions} or \code{man if} in the \code{spectrum} directory.

\subsubsection{Processes}
\label{subsub:proc}

The command \code{create random orbifold from(OrbifoldLabel)} starts a new child process that runs in the background. The commands referred to processes are \code{ps, kill(PID)} and \code{wait(X)}. The command \code{ps} lists the processes that are currently running in the \orbifolder{}. They are tagged with a PID number. The command \code{kill(PID)} kills the process associated with the PID number. The command \code{wait(X)} checks every \code{X} seconds if all processes have finished and to continue with the next commands afterwards. Additional details can be seen with the command \code{help processes} in any directory.

\subsubsection{Vectors}
\label{subsub:vect}

Some commands, like \code{set U1(i) = <16Dvector>} in the \code{gauge group} directory, need a 16D vector. The \orbifolder{} accepts several formats for these vectors. For example, the 16D vector

$(-18, 0, 6, 6, 0, 0, 0, -14, -38, 22, 0, 0, 0, 0, 0, 180)$

can also be defined as

$(-18\,\, 0\,\, 6\,\, 6\,\, 0\,\, 0\,\, 0\,\, -14\,\, -38\,\, 22\,\, 0\,\, 0\,\, 0\,\, 0\,\, 0\,\, 180)$

$(-18, 0, 6, 6, 0^3, -14, -38, 22, 0^5, 180)$

$(-18/1\,\, 0/1\,\, 6/1\,\, 6/1\,\, 0/1\,\, 0/1\,\, 0/1\,\, -14/1\,\, -38/1\,\, 22/1\,\, 0/1\,\, 0/1\,\, 0/1\,\, 0/1\,\, 0/1\,\, 180/1)$

$-18, 0, 6, 6, 0^3, -14, -38, 22, 0^5, 180$

These examples illustrate the different ways a 16D vector can be written. This also applies for the shift vectors and Wilson lines.

\subsubsection{Output for \LaTeX{} or in Mathematica style}
\label{subsub:outputml}

The output in \LaTeX{} or Mathematica style is available for some commands. For example, in the \code{spectrum} directory the commands \code{print summary}, \code{print summary of sectors} and \code{print(fields)}. In the \code{model} directory  the commands \code{print shift} and \code{print Wilson lines}. In the \code{gauge group} directory the command \code{print simple roots}. To activate the output in any of these two styles, add the parameters \code{@latex} or \code{@mathematica} after the name of the command. For example, to obtain the spectrum in a table with latex format, use the command \code{print summary @latex} in the \code{spectrum} directory.

To set the mode output in one style and change to another one use the commands \code{@typesetting(latex)}, \code{@typesetting(mathematica)} and \code{@typesetting(standard)}.

\subsubsection{System commands and variables}
\label{subsub:scv}

There are some system commands that change the output's style and destination. They start with the symbol @. Next, we present a brief description of them.

\paragraph{\code{@typesetting(Type)}}  Change the output to types: \LaTeX, Mathematica or standard. See~\ref{subsub:outputml}.

\paragraph{\code{@begin print to file(Filaname)}} It starts to print all outputs of the executed commands into a file named \code{Filename}. To print the output of only one command into a file use the parameter \code{to file(Filename)}. For example, the command \code{print summary to file(Filename)} in the \code{spectrum} directory.
%prints the output of the command \code{print summary} into a file named \code{Filename}.

\paragraph{\code{@end print to file}} It stops printing the outputs (that started with \code{@begin print to file(Filename)}) into \code{Filename}.

\paragraph{\code{@status}} It shows the current output and typesetting.\\

There are three pre-defined variables that are useful for scripts: \code{\$OrbifoldLabel\$}, \code{\$VEVConfigLabel\$} and \code{\$Directory\$}. When used they are replaced by the current orbifold label, vev-configuration and prompt directory, respectively. For example, suppose the \Z3 orbifold model with label \code{Z3\_1\_1} was loaded. Then, in the \code{spectrum} directory, the command \code{print summary to file(\$OrbifoldLabel\$.txt)} prints the corresponding output of the command \code{print summary} into a file named \code{Z3\_1\_1.txt}.

\subsection{The directories}
\label{sub:direct}

The prompt is structured in directories. The first two directories are the main directory and the orbifold model directory. For each orbifold model there are five directories: \code{model}, \code{gauge group}, \code{spectrum}, \code{vev-config}, and \code{vev-config/labels}. In this appendix we give an explanation of all the commands defined in these directories.

%main
\subsubsection{The main directory}
\label{subs:maindir}

This directory is identified by the symbol \code{>}. Here, the user can load, create, save, rename and delete orbifold models. The commands in this directory are:

\paragraph{\code{load orbifolds(Filename)}} This command loads orbifold(s) model(s) from a file named \code{Filename}. They are stored in orbifold directories with names equal to the orbifold labels. This command can be used with the parameter \code{inequivalent} to load only models with inequivalent massless spectra. The command \code{load orbifold(Filename)} works in a similar way.
\paragraph{\code{load program(Filename)}} This command loads a list of commands written, line by line, in a text file named \code{Filename}. The \orbifolder{} executes all these commands and shows their output in the prompt.
\paragraph{\code{save orbifolds(Filename)}} This command saves all orbifold models currently loaded in the \orbifolder{} (accessible in the main directory) to a file named \code{Filename}. The command \code{save orbifold(Filename)} works in a similar way.
\paragraph{\code{delete orbifold(OrbifoldLabel)}} This command deletes the orbifold model directory \code{OrbifoldLabel}.
\paragraph{\code{delete orbifolds}} This command deletes all orbifold model directories.
\paragraph{\code{rename orbifold(OldOrbifoldLabel) to(NewOrbifoldLabel)}} This command renames the orbifold label \code{OldOrbifoldLabel} to a new label \code{NewOrbifoldLabel}.
\paragraph{\code{create orbifold(OrbifoldLabel) with point group(M,N)}} This command creates an orbifold named \code{OrbifoldLabel} with point group \Z{M}\x\Z{N} (for \Z{M} set $N=1$). This point group is used for the compactification of the six additional spatial coordinates of the non-SUSY heterotic string on a 6D orbifold. It is understood, as we mentioned in section~\ref{sec:orbis}, that the complete point group is \Z{K}\x\Z{M}\x\Z{N} or \Z{K}\x\Z{M}, where $\Z{K} = \Z{2W}$ is the freely acting group that is used to construct the non-SUSY heterotic string from the SUSY heterotic string, where the twist vector $v_0$ of the $\Z{2W}$ group and its corresponding shift vector $V_0$ are fixed.

After executing this command, the \orbifolder{} creates a directory named \code{OrbifoldLabel}. When the user enters this directory, the \orbifolder{} asks for additional details like the space group geometry, the shift vectors $V_1$ and $V_2$ for a \Z{M}\x\Z{N} orbifold (or $V_1$ for a \Z{M} orbifold), and the Wilson lines $W_\alpha, \alpha = 1,\ldots,6$, to define completely the orbifold model.

\paragraph{\code{create orbifold(OrbifoldLabel) from(AnotherOrbifoldLabel)}} This command creates an orbifold model named \code{OrbifoldLabel} from another existing orbifold model called \code{AnotherOrbifoldLabel} (previously loaded or created). The new created model is equal to \code{AnotherOrbifoldLabel}. They differ in their names.
\paragraph{\code{create random orbifold from(OrbifoldLabel)}} This command creates randomly one orbifold model from another orbifold model previously loaded and labeled as \code{OrbifoldLabel}. This command can be complemented with several parameters. The user can type \code{help create random} to see them. They are:
\begin{itemize}
  \item \code{save to(Filename)}. It saves the orbifold(s) model(s) in a file named \code{Filename}.
  \item \code{if(...)}. It indicates the desired properties of the models. Use \code{inequivalent}  in order to create only models with inequivalent massless spectra, and use \code{SM}, \code{PS} or \code{SU5} for models with a net number of three generations {of fermions under the SM, PS or \SU5 gauge group} plus vector-like exotics. For example, \code{if(inequivalent SM)}.
  \item  \code{use(1,1,0,0,1,1,0,0)}. The first two digits are for to the two shifts, $V_1$ and $V_2$, corresponding to a \Z{M}\x\Z{N} orbifold, the remaining six digits are for the six Wilson lines $(W_1,...,W_6)$. The number 0 indicates that the shift or Wilson line, associated to the corresponding position of the 0 number, is created randomly, while the number 1 indicates that the shift or Wilson line is taken from the original orbifold model with label \code{OrbifoldLabel}.

  \item \code{\#models(X)}. It creates a number of \code{X} random models with the specified properties. To create as many as possible use \code{X = all}.
  \item \code{print info}. It prints a summary of the spectrum for the randomly created models.
  \item \code{load when done}. It loads the created models into orbifold directories after the process has finished.
  \item \code{do not check anomalies}. It speeds up the search and creation of the orbifold models with the specified properties.
\end{itemize}

%orbifold model directory
\subsubsection{The orbifold model directory}
\label{subs:orbimoddir}

This directory is identified as \code{/OrbifoldLabel>}, where \code{OrbifoldLabel} is the label of the orbifold model. From this directory the user can access the directories: {\code{model},\; \code{gauge group},\; \code{spectrum},\; \code{vev-config},\; and\; \code{vev-config/labels}}. The corresponding shortcuts {to access these directories} are \code{m}, \code{gg}, \code{s}, \code{v} and \code{l}. In these directories the user can explore different properties of the models. To enter one of these directories, for example, the \code{model} directory, type \code{cd model} or use the shortcut \code{m}. Similar steps apply to enter any of the other directories. A brief example illustrates the previous information. Consider the \Z3 orbifold model defined in the file \code{modelZ3\_1\_1.txt}. This model has the label \code{Z3\_1\_1}. Next, write
\begin{lstlisting}[style=consola,numbers=none]
$ ./orbifolder modelZ3_1_1.txt
\end{lstlisting}
\begin{lstlisting}[style=consola,numbers=none]
> cd Z3_1_1
\end{lstlisting}
\begin{lstlisting}[style=consola,numbers=none]
/Z3_1_1> cd model
\end{lstlisting}
\begin{lstlisting}[style=consola,numbers=none]
/Z3_1_1/model>
\end{lstlisting}
The orbifold model directory is identified as \code{/Z3\_1\_1>}.

%model
\subsubsection{The directory \code{model}}
\label{subs:modeldir}

This directory appears as \code{/model>} in the prompt. Here, the user can print and change input data for the orbifold model geometry. To see the commands in this directory type \code{/model> dir}. The \code{print} command allows for several options. Type \code{help print} to see them. The commands in this directory are:

\paragraph{\code{print orbifold label}} This command prints the orbifold label, which is also the name of the corresponding orbifold directory.

\paragraph{\code{print heterotic string type}} This command prints the 10D gauge group \SO{16}\x\SO{16} of the non-SUSY heterotic string.

\paragraph{\code{print available space groups}} This command  presents a list of the geometry files that are compatible with the orbifold point group. The geometry files are located in the directory \code{/localdirectory/Geometry>} of the local computer.

\paragraph{\code{print point group}} This command displays the point group of the orbifold model.

\paragraph{\code{print space group}} This command prints the point group, the root-lattice and the space group generators.

\paragraph{\code{print twist}} This command shows the twist vector(s) as 4D vector(s). The command \code{print twists} performs the same task.

\paragraph{\code{print \#SUSY}} This command prints the number of supersymmetry in 4D, which in this case is zero.

\paragraph{\code{print shift}} This command shows the shift(s) as 16D vectors. The command \code{print shifts} can also be used.

\paragraph{\code{print Wilson lines}} This command prints the relations among the Wilson lines, their order and the Wilson lines themselves as 16D vectors.

\paragraph{\code{use space group(i)}} This command loads the space group from the \code{i}-th geometry file. The list of the geometry files that are compatible with the orbifold point group appear with the command \code{print available space groups}.

\paragraph{\code{set shift V = <16D vector>}} This command sets the shift as a 16D vector for an orbifold model with point group \Z{M}. For the notation of 16D vectors see~\ref{subsub:vect}.

\paragraph{\code{set shift V(i) = <16D vector>}} This command sets the two shifts as 16D vectors for a \Z{M}\x\Z{N} orbifold model. Here \code{i}$ = 1,2.$

\paragraph{\code{set shift standard embedding}}  This command  sets the shift(s) in the standard embedding for the orbifold model. It indicates that the three first components of the shift vector $V=(V^1,V^2,V^3,0^5,0^8)$ are taken from the twist vector $v = (0,v^1,v^2,v^3)$ of a \Z{M} orbifold model such that $V^i=v^i$, for $i=1,2,3$. Similar assignations occur for the two shifts and two twists vectors in $\Z{M}\x\Z{N}$ orbifolds.

\paragraph{\code{set WL W(i) = <16D vector>}}  This command sets the \code{i}-th Wilson line as a 16D vector for the orbifold model. The index \code{i} takes values from 1 to 6.

%gauge group
\subsubsection{The directory \code{gauge group}}
\label{subs:ggdir}

This directory is identified as \code{/gauge group>} in the prompt. Here, the user can print details of the gauge group and change the \U1 basis. To see the commands in this directory type \code{/gauge group> dir}. The \code{print} command allows for several options. Type \code{help print} to see them. The commands in this directory are:

\paragraph{\code{print gauge group}}  This command prints the 4D gauge group in the current vev-configuration of the orbifold model.

\paragraph{\code{print beta coefficients}} This command computes the non-SUSY beta coefficients at one-loop for the non-Abelian gauge groups in the observable sector of the 4D gauge group.

\paragraph{\code{print simple roots}} This command prints a choice of simple roots as 16D vectors for the non-Abelian gauge groups. The number of simple roots corresponds to the rank of the non-Abelian gauge group factors contained in the 4D gauge group.

\paragraph{\code{print simple root(i)}} This command shows the \code{i}-th simple root as a 16D vector. The index \code{i} can be an integer number between 1 and \code{n}, where \code{n} is the total number of simple roots.

\paragraph{\code{print anomaly info}} This command displays information about the gauge and gravitational anomalies and verify their universality relations.

\paragraph{\code{print B-L generator}} This command prints the \code{B-L} generator as a 16D vector, which is introduced with the command \code{set B-L = <16D vector>}.

\paragraph{\code{print U1 generators}} This command shows all \U1 generators as 16D vectors.

\paragraph{\code{print U1 generator(i)}} This command prints the \code{i}-th \U1 generator as a 16D vector.

\paragraph{\code{set U1(i) = <16D vector>}} This command sets \U1 generators as 16D vectors. The index \code{i} indicates the \code{i}-th \U1 generator for an orbifold model. This assignation changes the basis of \U1 generators. The new generator must be orthogonal to all simple roots and to the \code{j}-th \U1 generator, for \code{j < i}. The \code{k}-th \U1 generators, for \code{k > i}, will be changed automatically, such that all generators are orthogonal to each other at the end. The anomalous \U1 cannot be changed. See section~\ref{subsub:vect} for details about the notation of 16D vectors.

\paragraph{\code{set B-L = <16D vector>}} This command defines $\U1_{\text{B-L}}$ as a 16D vector. \code{B-L} is stored as an additional vector because in the \orbifolder{} all \U1 generators are requested to be orthogonal to each other, however $\U1_{\text{B-L}}$ is in general not orthogonal to hypercharge. This command can be used with the parameter
\code{allow for anomalous B-L} if $\U1_{\text{B-L}}$ is allowed to mix with the anomalous \U1. See section~\ref{subsub:vect} for details about the notation of 16D vectors.

%spectrum
\subsubsection{The directory \code{spectrum}}
\label{subs:spectdir}

This directory is identified as \code{/spectrum>} in the prompt. Here, the user can print several details of the orbifold model spectrum. To see the commands in this directory type \code{/spectrum> dir}. An explanation of these commands is presented next.

\paragraph{\code{print summary}} This command shows the massless spectrum of the orbifold model along with their representations and charges under the observable sector of the 4D gauge group in the current vev-configuration. This command can be used with the following parameters (type \code{help print summary} to see them):

\begin{itemize}

  \item \code{with labels}. This command presents the spectrum with labels. For a given orbifold model and vev-configuration, fields of the 4D effective theory are referred with labels. For example, for a model in the vev-configuration \code{TestConfig1} the scalar and fermion fields are labeled as \code{s\_1, s\_2, ..., s\_n} and \code{f\_1, f\_2, ..., f\_m}, respectively. For a model in the vev-configuration \code{SMConfig1} the scalar and fermions fields are properly labeled. For example, labels as \code{h\_1, h\_2, ..., h\_n,} denote Higgs doublets and labels as \code{l\_1, l\_2, l\_3,} refer to left-handed lepton doublets.

  \item \code{of sectors}. This command shows the spectrum classified by the untwisted and twisted sectors where the scalar and fermion fields belong. The twisted sectors are denoted by $T(k,m,n)$, where $k,m,n$ are integer numbers. The untwisted sectors are indicated by $T(0,0,0)$, which appear as $U$ sector in the displayed information of this command.

  \item \code{of sector T(k,m,n)}. This command prints the spectrum for the specified sector $T(k,m,n)$. The sector $(k,m,n)$ refers to the twisted/untwisted sectors of the point group \Z{K}\x\Z{M}\x\Z{N} = \Z{2W}\x\Z{M}\x\Z{N}. As we mentioned in section~\ref{sec:orbis}, the \Z{2W} group is used in the construction of the non-SUSY heterotic string from the SUSY heterotic string and it is needed to specify all sectors when using this command.

Let us show some examples. For the \Z{3} orbifold the complete point group is \Z{2W}\x\Z{3} = \Z{K}\x\Z{M}, i.e. $K=2$ and $M=3$. A sector $T(k,m,n) = T(0,2,0)$ means the untwisted sector of \Z{2W} and the second twisted sector of \Z{3}. For the \Z{3}\x\Z{3} orbifold the complete point group is \Z{2W}\x\Z{3}\x\Z{3} = \Z{K}\x\Z{M}\x\Z{N}, i.e. $K=2$, $M=3$ and $N = 3$. Then, a sector $T(k,m,n) = T(1,2,1)$ refers to the twisted sector of \Z{2W}, the second twisted sector of \Z{3} and the first twisted sector of the second \Z{3}. Note that the sector $T(0,0,0)$ indicates the untwisted sector of \Z{2W}\x\Z{M}\x\Z{N} and \Z{2W}\x\Z{M} orbifolds. Recall that a cyclic group of order $M$ is defined as $\Z{M} = \{\theta^m\,\vert\, \theta^{0\,\text{mod}\,M} = \Id\}$, where $m = 0,1,2,3,...,M-1$. The element $\theta$ is the generator of this group. The sector associated to $m = 0$ is the untwisted sector, and the sectors corresponding to $m = 1,2,3,...,M-1$ are the twisted sectors.

  \item \code{of fixed points}. This command presents the sector $(k,m,n)$, the label for the fixed point and six integer numbers $(n_1,n_2,n_3,n_4,n_5,n_6)$ of the translational part of the space group element associated to the fixed point, the 16D localization vector $V_{loc}$, and the field representation under the 4D gauge group in the current vev-configuration.
If some sector does not contain particle fields, then the word empty appears instead of the representation.

The notation $(k,m,n) (n_1,n_2,n_3,n_4,n_5,n_6)$ refers to the space group element $g=(\beta^k \theta^m \omega^n, n_1 e_1 + n_2 e_2 + ... + n_6 e_6)$, where $\beta$, $\theta$ and $\omega$ are the generators of the group factors in the point group $\Z{K}\x\Z{M}\x\Z{N} = \Z{2W}\x\Z{M}\x\Z{N}$, respectively. Then, $k=0,1$, $m = 0,1,2,\ldots,M-1$, and $n = 0,1,2,\ldots, N-1$. The set of integer numbers $(n_1,n_2,...,n_6)$ indicates $n_1 e_1 + n_2 e_2 + ... + n_6 e_6 = n_\alpha e_\alpha,\,\alpha = 1,...,6$, where $e_\alpha$ are the 6D torus lattice basis vectors. As we mentioned in section~\ref{sec:orbis}, each fixed point has a corresponding space group element $g=(\beta^k \theta^m \omega^n, n_\alpha e_\alpha)$ called the constructing element. Then, a fixed point can be specified by the set of numbers $(k,m,n)(n_1,n_2,...,n_6)$. The part $(\beta^k \theta^m \omega^n)$ is the rotational part of the space group element and it is used to specify the untwisted and twisted sectors by the set of numbers $(k,m,n)$. The linear combination $n_1 e_1 + .... + n_6 e_6$ is the translational part of the space group element. In the case of roto-translations, the numbers $n_\alpha$ are not integers.

  \item \code{of fixed point(label)}. It prints the same information as the previous command but only for the fixed point with the specified label, which can be seen with the previous command \code{print summary of fixed points}.

  \item \code{of fixed point(k,m,n,n1,n2,n3,n4,n5,n6)}. It displays the same details as \code{print summary of fixed point(label)} but now by specifying the sector $(k,m,n)$ and the numbers $n_a = (n_1,n_2,...,n_6)$ of the fixed point instead of the label. Recall that the label, the sector $(k,m,n)$ and the numbers $(n_1,n_2,n_3,n_4,n_5,n_6)$ associated to a fixed point are provided with the command \code{print summary of fixed points}.

  \item \code{no U1s}. It shows the spectrum without the \U1 charges.

The parameters \code{no U1s} and \code{with labels} can be used together with the command \code{print summary} and the other parameters. For example, \code{print summary no U1s with labels}, \code{print summary of sectors no U1s with labels}, etc.
\end{itemize}

\paragraph{\code{print(fields)}} For a specified field label this command shows the sector $(k,m,n)$ of the $\Z{2W}\x\Z{M}\x\Z{N}$ point group, the numbers
$(n_1,n_2,...,n_6)$ of the translational part of the space group element, the representation of the field under the 4D gauge group in the current vev-configuration, the left-moving momenta, the right-moving momentum, and the oscillators acting on left states.
Recall that for a given orbifold model and vev-configuration, fields of the 4D effective theory are referred to with labels, which can be seen with the command \code{print summary with labels}. The word \code{fields} inside the parentheses of \code{print(fields)} refers to the label of a field or a set of labels for fields. For instance, \code{print(\text{s\_7})} displays the respective information for the scalar field labeled as s\_7, while \code{print(\text{f\_1 s\_7})} presents the details for the fermion field f\_1 and the scalar field s\_7. Use \code{print(*)} to access all fields in the spectrum. See~\ref{subsub:fieldl} for details about field labels.\\
\noindent The command \code{print(fields)} can be used with the parameter

\code{with internal information}

\noindent In this case additional details for the fields such as the gamma phases, \code{internalIndex} and field number are also printed. They represent internal information about how the fields' data can be accessed in the \code{C++} source code of the \orbifolder{}.

\paragraph{\code{print all states}} For all fields in the spectrum of an orbifold model this command presents: the untwisted and twisted sectors $(k,m,n)$, the numbers $(n_1,n_2,...,n_6)$ of the translational part of the constructing element, the label of the fixed point, the representation of the field under the 4D gauge group in the current vev-configuration, the field label, the oscillators acting on left states, the left-moving momenta, the right-moving momentum and the gamma phases.

\paragraph{\code{print list of charges(fields)}} This command prints the left-moving momenta and the right-moving momentum of \code{fields} specified by their labels. For example, \code{print list of charges(\text{s\_5})}. To consider all fields in the spectrum use \code{print list of charges(*)}. The command can be used with the parameter

\code{label of list(Label)}

\noindent In this case the information displayed from \code{print list of charges(fields)} is tagged as \code{Label}.

\paragraph{\code{tex table(fields)}}  This command prints a table in \LaTeX{} format for the spectrum of the observable sector in the current vev-configuration. The word \code{fields} refers to the field label. For example,
\code{tex table(f\_7)} for a fermion field labeled as \code{f\_7}. For all fields in the spectrum use \code{*} instead of \code{fields}, i.e.\ \code{tex table(*)}. This command can be complemented with the parameter

\code{print labels(i)}

\noindent where \code{i} indicates the \code{i}-th labeling for the fields, which can be seen in the \code{vev-config/labels} directory with the command \code{print labels} (see~\ref{subs:labeldir}). For example, \code{tex table(*) print labels(1)} prints a \LaTeX{} table for the scalar and fermion fields classified by untwisted and twisted sectors, it also provides the fields representations under the 4D gauge group and the field labels associated to the \code{i}-th labeling, where \code{i=1} in this example.

This command can be used with the parameter \code{to file(Filename)} to get the output of the command in a file named \code{Filename} (see~\ref{subsub:scv}). For example, \code{tex table(*) print labels(1) to file(textable.tex)} prints the output of the command \code{tex table(*) print labels(1)} in a tex file named \code{textable.tex}.

%vev-config
\subsubsection{The directory \code{vev-config}}
\label{subs:vdir}

This directory is identified as \code{/vev-config>}. Here, the user can define and analyse the vev-configurations of the orbifold model and select its observable sector. To see the commands type \code{/vev-config> dir}. The commands in this directory are:

\paragraph{\code{use config(ConfigLabel)}} This command changes the currently used vev-configuration of an orbifold model to another existing vev-configuration with name \code{ConfigLabel}.

\paragraph{\code{create config(ConfigLabel)}} This command creates a new vev-configuration named \code{ConfigLabel}. Its origin is the standard vev-configuration named \code{StandardConfig1}. \\
This command can be used with the parameter

\code{from(AnotherConfigLabel)}\\
It creates a new vev-configuration named \code{ConfigLabel} from another defined vev-configuration named \code{AnotherConfigLabel}.

\paragraph{\code{rename config(OldConfigLabel) to(NewConfigLabel)}} This command renames a vev-configuration with name \code{OldConfigLabel} to a new name given by \code{NewConfigLabel}.

\paragraph{\code{delete config(ConfigLabel)}} This command deletes a vev-configuration named \code{ConfigLabel}.

\paragraph{\code{print configs}} This command prints a list of vev-configurations defined for the orbifold model. The currently vev-configuration is indicated by an arrow.

\paragraph{\code{print gauge group}} This command prints the gauge group for the selected choice of observable and hidden sector of the currently used vev-configuration. Gauge group factors that belong to the hidden group are in brackets.

\paragraph{\code{select observable sector:\,[parameters]}} This command allows to select different gauge group factors as part of the observable sector from a gauge group in the current vev-configuration of an orbifold model. The possible parameters are:
\begin{itemize}
  \item\code{gauge group(i,j,...)}. The indices \code{i,j} enumerate the position of the non-Abelian gauge group factors that form the gauge group. The position of these groups can be noticed by using the command \code{print gauge group}. The selected non-Abelian gauge groups are part of the observable sector.

  \item \code{full gauge group}. All non-Abelian gauge group factors are chosen as part of the observable sector.

  \item \code{no gauge groups}. None of the non-Abelian gauge group factors are part of the observable sector, i.e.\ all of them belong to the hidden sector.

  \item \code{U1s(i,j,...)}. The indices \code{i,j} enumerate the Abelian \U1 gauge factors. They can be seen with the command \code{print gauge group}. The chosen Abelian gauge groups are part of the observable sector.

  \item \code{all U1s}. All Abelian \U1 factors are part of the observable sector.

  \item \code{no U1s}. None of the Abelian \U1 factors form part of the observable sector, i.e.\ all of them belong to the hidden sector.
\end{itemize}

For example, the gauge group of the \Z3 orbifold model (defined in the file \code{modelZ3\_1\_1.txt}) is
$\SO{10}\x\SU{3}\x\SO{16}\x\U1$. Then, the command \code{select observable sector: gauge group(1,3) U1s(1)}
selects $\SO{10}\x\SO{16}\x\U1$ as the observable sector, and \SU{3} builds the hidden sector.

\paragraph{\code{analyze config}} This command checks if the current orbifold model allows for vacua with SM, PS or \SU5 gauge group, three generations of fermions and vector-like exotics. If one of these possibilities is realised then the corresponding spectrum with appropriate field labels is printed. This command can be used with the parameter

\code{print SU(5) simple roots}\\
In this case the simple roots of an intermediate \SU{5} gauge group that has been used to identify the hypercharge generator are also displayed.

%vev-config/label
\subsubsection{The directory \code{vev-config/labels}}
\label{subs:labeldir}

This directory is identified as \code{/vev-config/labels>}. Here, the user can assign labels for the fields of the massless spectrum. Write \code{dir}, i.e.\ \code{/vev-config/labels> dir}, to see the commands in this directory. Next, we give a brief description of them.

\paragraph{\code{change label (A\_i) to(B\_j)}} This command changes the label of the field \code{A\_i} to \code{B\_j}.

\paragraph{\code{create labels}} This command shows the massless spectrum, then the user is asked to write labels for each line in the spectrum. This task is first performed for the scalars and then for the fermions.

\paragraph{\code{assign label(Label) to fixed point(k,m,n,n1,n2,n3,n4,n5,n6)}} This command assigns the label \code{Label} to the fixed point with localization \code{(k,m,n,n1,n2,n3,n4,n5,n6)}.

\paragraph{\code{print labels}} This command prints the \code{i}-th labeling and the massless spectrum with the corresponding labels for the scalar and fermion fields.

\paragraph{\code{use label(i)}} This command changes the currently used labels to the \code{i}-th labeling.

\paragraph{\code{save labels(Filename)}} This command saves the currently \code{i}-th labeling to a file named \code{Filename}.

\paragraph{\code{load labels(Filename)}} This command loads the labels from \code{Filename} and shows the massless spectrum with the corresponding field labels.

%man help
\subsection{The \code{man} utility}
\label{sub:man}

In each directory of the \orbifolder{}, one can use the command \code{man} to get manuals for the available commands as well as several examples and some details. To illustrate how the command \code{man} works, let us consider the main directory, identified with the symbol  \code{>}. Type \code{man} to see a list of short command names that can be used with the command \code{man} as \code{man name}, where \code{name} is a short command name. In the main directory these short command names are: \code{cd, create, delete, load, rename and save}. Suppose the user wants to know some details and examples for \code{create}. Then, write
\begin{lstlisting}[style=consola,numbers=none]
> man create
\end{lstlisting}
This will open a new screen on the terminal showing information organized in sections. Some of them are identified as: {\footnotesize{NAME, SYNOPSIS, DESCRIPTION, OPTIONS}}, and {\footnotesize{EXAMPLES}}. The {\footnotesize{OPTIONS}} section presents different possibilities for the \code{create} command. For example,
\begin{itemize}
\item \code{orbifold(OrbifoldLabel) with point group(M,N)}
\item \code{orbifold(OrbifoldLabel) from(AnotherOrbifoldLabel)}\item \code{random orbifold from(OrbifoldLabel)}
\end{itemize}
and possible additional parameters that are defined for some commands. There are cases where the full command name is the only option. For example, when typing \code{> man rename}, the only option for the \code{rename} command is \code{orbifold(OldOrbifoldLabel) to(NewOrbifoldLabel)}. The {\footnotesize{EXAMPLES}} section shows several useful examples that could help the user be familiar with the use of these commands and understand some notation. The use of the command \code{man} in all other directories is analogous.

\newpage
% ------------------------------------------------------------

\section{Using the non-SUSY Orbifolder via a Docker Container}
\label{app:docker}

\subsection{Docker Functionality}
An additional feature introduced in the \orbifolder{} is its compatibility with operating systems beyond those based on GNU/Linux, including Windows 10 and macOS Sequoia. In order to facilitate the execution of the \orbifolder{} on these platforms, it is necessary to employ a software packaging and virtualization solution known as Docker.
This technology enables the deployment and execution of applications originally designed for a specific operating system within environments that differ from the original system.
This interoperability is achieved through the use of \emph{containers}, which constitute the core abstraction of the Docker platform, providing isolated and consistent runtime environments across heterogeneous systems.

\begin{figure}[b!]
    \centering
    \includegraphics[scale=0.4]{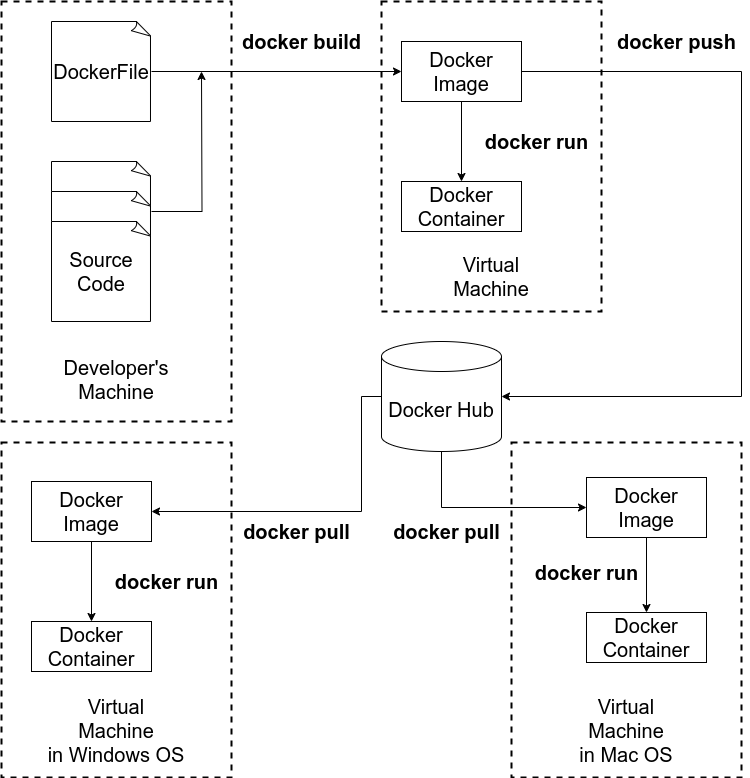}
    \caption{Docker workflow: The implementation
    of Docker technology commence with the software operating natively on a host operating system.
    Concurrently, it is necessary to define a configuration file, known as the \texttt{DockerFile},
    which defines the base system (e.g.\ operating system name and version),
    as well as the essential libraries required to ensure the proper
    functioning of the software to be deployed.
    Subsequently, a \texttt{Docker Image} is created. This image constitutes a portable and
    reproducible snapshot of the configured environment, including the operating system and
    the software components. The resulting image may be distributed via the Docker Hub
    platform or through local repositories, depending on the intended deployment strategy.
    On the target system where the \texttt{non-SUSY orbifolder} application is to be executed,
    access to the image-whether retrieved from Docker Hub or a local source is
    required. From this image, a \texttt{container} is instantiated, which represents an
    isolated and self-sufficient execution environment for the software.
    Once the container is instantiated, the execution of commands and the behavior of the system
    remain consistent, regardless of the underlying host system, thereby ensuring portability
    and reliability across diverse computing environments.}
    \label{fig:dockerarchitecture}
\end{figure}

Figure \ref{fig:dockerarchitecture} illustrates the Docker workflow, from the creation
of a container by a developer to its deployment in different environments. This process
begins with the developer, who creates a file called DockerFile. This file contains a
series of instructions that describe how to build a Docker image. These instructions
specify the environment configuration,
including the required dependencies, software installation, port and volume configurations,
and the files to be included in the container. The DockerFile is essentially the template
from which images are constructed.

From the provided DockerFile, a Docker image is constructed. This image constitutes a static
and portable snapshot of the complete runtime environment, encapsulating all components required
for the proper execution of the application, including the base operating system, necessary
libraries, configuration files, and the application code itself. Once built, the image can be
executed on any system with Docker installed, ensuring a reproducible and isolated environment
across different platforms. Executing the image instantiates a container,
which represents the active, runtime manifestation of the image. The container provides a dynamic
environment through which users can interact with the encapsulated application.

The workflow proceeds with the uploading the generated image to Docker Hub,
a centralized repository designed to streamline the distribution and sharing of Docker
images. Once hosted on Docker Hub, the image can be retrieved by other systems through a
process referred to as a pull. On these target systems,
the image is instantiated once again as a container.

In summary, the workflow illustrated in Figure~\ref{fig:dockerarchitecture}
exemplifies how Docker streamlines the creation, distribution, and utilization
of consistent and reproducible runtime environments across all stages of the software
development life cycle. From the developer's local environment to target systems,
Docker ensures environmental consistency, thereby minimizing
compatibility issues and enhancing the overall efficiency and reliability of
application deployment.

The DockFile used to construct the image of the \orbifolder{} is defined as follows
\begin{lstlisting}[language=Dockerfile]
FROM ubuntu:22.04
RUN yes | unminimize
# Libraries
RUN apt-get update && apt-get install -y \
    build-essential \
    man-db \
    less \
    g++ \
    cmake \
    libgsl-dev \
    nano \
    libboost-math-dev \
    libreadline-dev
# Setup work directory
WORKDIR /app
# Copy file to container
COPY . .
# Compilation
RUN ./configure
RUN make
RUN make install
# Remove unnecessary files
#RUN rm -rf /var/lib/apt/lists/*
# Execute the program
CMD ["./nonSUSYorbifolder"]
\end{lstlisting}

It is important to note that the base system employed for the development of this
new version was Ubuntu 22.04. Nevertheless, as previously
indicated, its functionality was also successfully tested on other GNU/Linux
distributions, including Ubuntu 16.04, {18.04, 20.04, and} 24.04, Linux
Mint 21, and Fedora 39. The resulting Docker image has a size of approximately
1.52GB. However, this increase in size is justified by the substantial time
savings achieved during both development and deployment phases, as Docker
provides a seamless and efficient mechanism for running the \orbifolder{}
on alternative operating systems, such as Windows~10.

\subsection{Docker Installation}

Installing Docker is a relatively straightforward process that enables
the configuration of this tool on various operating systems for efficient
container management. Docker is available across multiple platforms, including
Linux, Windows, and macOS. While the specific installation procedures may vary
slightly depending on the operating system, the underlying principles remain
consistent across all supported environments

As a preliminary step, it is essential to ensure that the system meets the
necessary prerequisites. For Linux-based distributions such as Ubuntu or CentOS,
this involves having superuser (root) privileges or access to the \code{sudo} command,
as well as an up-to-date system with the latest package versions. In contrast,
on Windows and macOS platforms, the installation process is facilitated through
Docker Desktop (an integrated solution that streamlines setup and management).
However, its proper functioning requires additional system capabilities,
such as virtualization support enabled at the BIOS or firmware level.

The subsequent step involves downloading and installing Docker from the official
repositories. On Linux-based systems, this process generally consists of adding
the Docker repository and utilizing the native package manager such
as \texttt{apt} for Ubuntu or \texttt{yum} for CentOS. A typical Ubuntu
installation procedure begins with updating the system package index
\begin{lstlisting}[style=consola,numbers=none]
$ sudo apt update
\end{lstlisting}
This is followed by the installation of the necessary dependencies, the addition
of Docker's official GPG key to ensure package authenticity, and the inclusion
of the Docker repository in the system sources list. Once these steps are
completed, the Docker Engine package can be installed using
\begin{lstlisting}[style=consola,numbers=none]
$ sudo apt install docker-ce
\end{lstlisting}

For Windows and macOS, Docker Desktop can be downloaded directly from Docker
official website, and it includes a graphical installer that guides the user
through the process.

Once the installation is complete, it is important to verify that Docker has been installed
correctly. This is done by running the command
\begin{lstlisting}[style=consola,numbers=none]
$ docker --version
\end{lstlisting}
to check the installed version of Docker, or by executing
\begin{lstlisting}[style=consola,numbers=none]
$ docker run hello-world
\end{lstlisting}
which performs a simple test to ensure that the Docker service is functioning properly.
If the service is not active, it may need to be started manually using
\begin{lstlisting}[style=consola,numbers=none]
$ sudo systemctl start docker
\end{lstlisting}

Finally, to simplify the use of Docker, it is recommended to add the current user to the Docker
group, which removes the need superuser permissions
\begin{lstlisting}[style=consola,numbers=none]
$ sudo usermod -aG docker $USER
\end{lstlisting}
followed by logging out and back in to apply the changes.

\subsection{Image Downloading}

One of functionalities of Docker is the ability to retrieve images from Docker Hub,
a centralized repository designed to streamline the acquisition and distribution of
preconfigured environments. These images serve as foundational layers upon which applications
can be built and executed. Docker Hub allows developers to store and share images, offering
both public and private access options depending on the intended use case and access control
requirements.

When a user wants to download an image from Docker Hub, the process begins with the
\texttt{docker pull} command. This command allows specifying the name of the image to be downloaded.
In its simplest form, the command takes the image name as an argument, which can include a username
or an organization if the image is stored in a private or specific repository. For example,
\texttt{docker pull ubuntu} will download the latest official version of the Ubuntu image.

Docker also allows specifying tags to download a specific version of an image. Each image in
Docker Hub can have multiple tags representing different versions or configurations. For instance,
running \texttt{docker pull nginx:alpine} will download a lightweight version of Nginx based on
Alpine Linux, whereas \texttt{docker pull nginx:latest} will download the most recent version of
the Nginx web server. If no tag is specified, Docker uses the \emph{latest} tag by default.

When the \texttt{docker pull} command is executed, Docker first checks whether the image already exists
locally. If the image is available and no updates are detected, Docker takes no
further action. However, if the image is not present or a newer version exists on Docker Hub,
Docker starts the download process. This involves transferring the image from Docker Hub servers
to the local system. Each Docker image consists of multiple layers, and Docker optimizes the
process by downloading only the layers that are not already available locally.

Once all layers have been downloaded, Docker assembles them to form the complete image. The image
is stored locally and becomes ready for use. Users can verify which images are available on their
system
\begin{lstlisting}[style=consola,numbers=none]
$ docker images
\end{lstlisting}
this command displays a list of all images stored locally, along with their tags, sizes, and unique IDs.

In environments with access restrictions, such as when downloading images from private repositories
on Docker Hub, the user must log in beforehand
\begin{lstlisting}[style=consola,numbers=none]
$ docker login
\end{lstlisting}
this ensures Docker has the necessary credentials to access the specified repository.

To download the \orbifolder{} image, execute the command
\begin{lstlisting}[style=consola,numbers=none]
$ docker pull stringsifunam/nonsusyorbifolder:v1
\end{lstlisting}

\subsection{Container Creation}

Creating containers interactively in Docker is a useful technique when it is necessary to
work directly within the container's environment, whether for configuration, debugging,
or immediate testing. Unlike running containers in the background, interactive mode allows
the user to have an active session in the container's command line, providing an experience
similar to working on a lightweight virtual machine.

To create a container interactively, the main command is \texttt{docker run} accompanied by
the \texttt{-it} options. The \texttt{-i} option (interactive) keeps the standard input
(stdin) open, while the \texttt{-t} option assigns a pseudo-terminal to make the interaction
experience smoother. By combining both options, the user can interact with the container
in real-time, entering commands and receiving responses directly in the console.
Additionally, the \texttt{--rm} option can be used to automatically remove the container
once it is stopped, which is useful for temporary containers that do not need to persist
after use. To launch the container in interactive mode, execute
\begin{lstlisting}[style=consola,numbers=none]
$ docker run --name nonsusyorbifoldercont --rm -it stringsifunam/nonsusyorbifolder:v1 bash
\end{lstlisting}

In this scenario, Docker creates a container based on Ubuntu and grants the user direct access
to a command console inside the container. Once inside, the user may run the
\orbifolder{} in the same way as explained in section~\ref{sec:structure}.

It is important to note that if the container stops or the session is closed, all modifications
made in that environment will be lost unless action is taken to save them. To preserve changes,
the container can be committed to a new image using the \emph{commit} command. This allows saving
the current state of the container as a reusable image, in other terminal of the host system typing
\begin{lstlisting}[style=consola,numbers=none]
$ docker commit <container_id> nonsusyorbifolder:mytag
\end{lstlisting}

The \texttt{container\_id} can be obtained from the information displayed by the execution of
\begin{lstlisting}[style=consola,numbers=none]
$  docker ps
\end{lstlisting}
which shows all containers currently running.

Additionally, if you want to link directories or files from the host system to the container
while working interactively, the \texttt{-v} option can be used to mount volumes. For example,
\begin{lstlisting}[style=consola,numbers=none]
$ docker run -it -v /path/local:/app stringsifunam/nonsusyorbifolder:v1 bash
\end{lstlisting}
This allows a directory from the host system to be accessible inside the container, facilitating
data exchange between both environments.

To exit an interactive container, you can use the \texttt{exit} command, which will stop the
container by default. If you wish to exit but keep the container running, the key combination
\texttt{Ctrl+P + Ctrl+Q} can be used, which detaches the session without terminating the container.

Finally, it is possible to remove images stored on the host system using
\begin{lstlisting}[style=consola,numbers=none]
$ docker rmi image_id
\end{lstlisting}

%%%%%%%%%%%%%%%%%%%%%%%%%%%%%%%%%%%%%%%%%%%%%%%%%%%%%%%%%%%%%%%%%%%%%%%%%%
%  Bibliography
%%%%%%%%%%%%%%%%%%%%%%%%%%%%%%%%%%%%%%%%%%%%%%%%%%%%%%%%%%%%%%%%%%%%%%%%%%
%\clearpage
%\newpage
%\bibliographystyle{OurBibTeX}
%\bibliography{Orbifold}

\begin{thebibliography}{10}

\bibitem{Gross:1984dd}
D.~J. Gross, J.~A. Harvey, E.~J. Martinec, and R.~Rohm, \emph{The heterotic
  string}, Phys. Rev. Lett. \textbf{54} (1985), 502--505.

\bibitem{Dixon:1986iz}
L.~J. Dixon and J.~A. Harvey, \emph{{String Theories in Ten-Dimensions Without
  Space-Time Supersymmetry}}, Nucl. Phys. B \textbf{274} (1986), 93--105.

\bibitem{Alvarez-Gaume:1986ghj}
L.~\'Alvarez-Gaum\'e, P.~H. Ginsparg, G.~W. Moore, and C.~Vafa, \emph{{An O(16)
  x O(16) Heterotic String}}, Phys. Lett. B \textbf{171} (1986), 155--162.

\bibitem{Baykara:2024tjr}
Z.~K. Baykara, H.-C. Tarazi, and C.~Vafa, \emph{{New Non-Supersymmetric
  Tachyon-Free Strings}},  (2024), \texttt{arXiv:2406.00185} [hep-th].

\bibitem{Larotonda:2024thv}
V.~Larotonda and L.~Lin, \emph{{Anomaly Inflow and Gauge Group Topology in the
  10d Sugimoto String Theory}},  (2024), \texttt{arXiv:2412.17894} [hep-th].

\bibitem{Basile:2023knk}
I.~Basile, A.~Debray, M.~Delgado, and M.~Montero, \emph{{Global anomalies \&
  bordism of non-supersymmetric strings}}, JHEP \textbf{02} (2024), 092,
  \texttt{arXiv:2310.06895} [hep-th].

\bibitem{Abel:2015oxa}
S.~Abel, K.~R. Dienes, and E.~Mavroudi, \emph{{Towards a nonsupersymmetric
  string phenomenology}}, Phys. Rev. D \textbf{91} (2015), no.~12, 126014,
  \texttt{arXiv:1502.03087} [hep-th].

\bibitem{Ashfaque:2015vta}
J.~M. Ashfaque, P.~Athanasopoulos, A.~E. Faraggi, and H.~Sonmez,
  \emph{{Non-Tachyonic Semi-Realistic Non-Supersymmetric Heterotic String
  Vacua}}, Eur. Phys. J. C \textbf{76} (2016), no.~4, 208,
  \texttt{arXiv:1506.03114} [hep-th].

\bibitem{Blaszczyk:2015zta}
M.~Blaszczyk, S.~Groot~Nibbelink, O.~Loukas, and F.~Ruehle, \emph{{Calabi-Yau
  compactifications of non-supersymmetric heterotic string theory}}, JHEP
  \textbf{10} (2015), 166, \texttt{arXiv:1507.06147} [hep-th].

\bibitem{Abel:2017vos}
S.~Abel, K.~R. Dienes, and E.~Mavroudi, \emph{{GUT precursors and entwined
  SUSY: The phenomenology of stable nonsupersymmetric strings}}, Phys. Rev. D
  \textbf{97} (2018), no.~12, 126017, \texttt{arXiv:1712.06894} [hep-ph].

\bibitem{Faraggi:2020fwg}
A.~E. Faraggi, V.~G. Matyas, and B.~Percival, \emph{{Type 0
  \ensuremath{\mathbb{Z}}2 \texttimes{} \ensuremath{\mathbb{Z}}2 heterotic
  string orbifolds and misaligned supersymmetry}}, Int. J. Mod. Phys. A
  \textbf{36} (2021), no.~24, 2150174, \texttt{arXiv:2010.06637} [hep-th].

\bibitem{Aoyama:2020aaw}
K.~Aoyama and Y.~Sugawara, \emph{{Non-SUSY Gepner Models with Vanishing
  Cosmological Constant}}, PTEP \textbf{2020} (2020), no.~10, 103B01,
  \texttt{arXiv:2005.13198} [hep-th].

\bibitem{Aoyama:2021kqa}
K.~Aoyama and Y.~Sugawara, \emph{{Non-SUSY Heterotic String Vacua of Gepner
  Models with Vanishing Cosmological Constant}}, PTEP \textbf{2021} (2021),
  no.~3, 033B03, \texttt{arXiv:2102.00683} [hep-th].

\bibitem{Faraggi:2020wld}
A.~E. Faraggi, V.~G. Matyas, and B.~Percival, \emph{{Classification of
  nonsupersymmetric Pati-Salam heterotic string models}}, Phys. Rev. D
  \textbf{104} (2021), no.~4, 046002, \texttt{arXiv:2011.04113} [hep-th].

\bibitem{Faraggi:2020hpy}
A.~E. Faraggi, V.~G. Matyas, and B.~Percival, \emph{{Type $\mathbf{\bar{0}}$
  heterotic string orbifolds}}, Phys. Lett. B \textbf{814} (2021), 136080,
  \texttt{arXiv:2011.12630} [hep-th].

\bibitem{Florakis:2021bws}
I.~Florakis, J.~Rizos, and K.~Violaris-Gountonis, \emph{{Super no-scale models
  with Pati-Salam gauge group}}, Nucl. Phys. B \textbf{976} (2022), 115689,
  \texttt{arXiv:2110.06752} [hep-th].

\bibitem{Faraggi:2022hut}
A.~E. Faraggi, V.~G. Matyas, and B.~Percival, \emph{{Towards classification of
  N=1 and N=0 flipped SU(5) asymmetric Z2\texttimes{}Z2 heterotic string
  orbifolds}}, Phys. Rev. D \textbf{106} (2022), no.~2, 026011,
  \texttt{arXiv:2202.04507} [hep-th].

\bibitem{Blaszczyk:2014qoa}
M.~Blaszczyk, S.~Groot~Nibbelink, O.~Loukas, and S.~Ramos-S\'anchez,
  \emph{{Non-supersymmetric heterotic model building}}, JHEP \textbf{10}
  (2014), 119, \texttt{arXiv:1407.6362} [hep-th].

\bibitem{Perez-Martinez:2021zjj}
R.~P\'erez-Mart\'inez, S.~Ramos-S\'anchez, and P.~K.~S. Vaudrevange,
  \emph{{Landscape of promising nonsupersymmetric string models}}, Phys. Rev. D
  \textbf{104} (2021), no.~4, 046026, \texttt{arXiv:2105.03460} [hep-th].

\bibitem{Dixon:1985jw}
L.~J. Dixon, J.~A. Harvey, C.~Vafa, and E.~Witten, \emph{{Strings on
  Orbifolds}}, Nucl. Phys. B \textbf{261} (1985), 678--686.

\bibitem{Dixon:1986jc}
L.~J. Dixon, J.~A. Harvey, C.~Vafa, and E.~Witten, \emph{{Strings on Orbifolds.
  2.}}, Nucl. Phys. B \textbf{274} (1986), 285--314.

\bibitem{Bailin:1999nk}
D.~Bailin and A.~Love, \emph{Orbifold compactifications of string theory},
  Phys. Rept. \textbf{315} (1999), 285--408.

\bibitem{Ramos-Sanchez:2008nwx}
S.~Ramos-S{\'a}nchez, \emph{{Towards Low Energy Physics from the Heterotic
  String}}, Fortsch. Phys. \textbf{57} (2009), 907--1036,
  \texttt{arXiv:0812.3560} [hep-th].

\bibitem{Vaudrevange:2008sm}
P.~K.~S. Vaudrevange, \emph{{Grand Unification in the Heterotic Brane World}},
  (2008), \texttt{arXiv:0812.3503} [hep-th].

\bibitem{Ramos-Sanchez:2024keh}
S.~Ramos-S\'anchez and M.~Ratz, {Heterotic Orbifold Models}, Springer, 2024.

\bibitem{Sagnotti:1996qj}
A.~Sagnotti, \emph{{Surprises in open string perturbation theory}}, Nucl. Phys.
  B Proc. Suppl. \textbf{56} (1997), 332--343, \texttt{hep-th/9702093}.

\bibitem{Angelantonj:1998gj}
C.~Angelantonj, \emph{{Nontachyonic open descendants of the 0B string theory}},
  Phys. Lett. B \textbf{444} (1998), 309--317, \texttt{hep-th/9810214}.

\bibitem{Blumenhagen:1999ns}
R.~Blumenhagen, A.~Font, and D.~Lust, \emph{{Tachyon free orientifolds of type
  0B strings in various dimensions}}, Nucl. Phys. B \textbf{558} (1999),
  159--177, \texttt{hep-th/9904069}.

\bibitem{Moriyama:2001ge}
S.~Moriyama, \emph{{USp(32) string as spontaneously supersymmetry broken
  theory}}, Phys. Lett. B \textbf{522} (2001), 177--180,
  \texttt{hep-th/0107203}.

\bibitem{Gato-Rivera:2007ifz}
B.~Gato-Rivera and A.~N. Schellekens, \emph{{Non-supersymmetric Tachyon-free
  Type-II and Type-I Closed Strings from RCFT}}, Phys. Lett. B \textbf{656}
  (2007), 127--131, \texttt{arXiv:0709.1426} [hep-th].

\bibitem{Fischer:2012qj}
M.~Fischer, M.~Ratz, J.~Torrado, and P.~K.~S. Vaudrevange,
  \emph{{Classification of symmetric toroidal orbifolds}}, JHEP \textbf{01}
  (2013), 084, \texttt{arXiv:1209.3906} [hep-th].

\bibitem{GrootNibbelink:2017luf}
S.~Groot~Nibbelink, O.~Loukas, A.~M\"utter, E.~Parr, and P.~K.~S. Vaudrevange,
  \emph{{Tension Between a Vanishing Cosmological Constant and
  Non-Supersymmetric Heterotic Orbifolds}}, Fortsch. Phys. \textbf{68} (2020),
  no.~7, 2000044, \texttt{arXiv:1710.09237} [hep-th].

\bibitem{Ploger:2007iq}
F.~Pl{\"o}ger, S.~Ramos-S{\'a}nchez, M.~Ratz, and P.~K.~S. Vaudrevange,
  \emph{{Mirage Torsion}}, JHEP \textbf{04} (2007), 063,
  \texttt{hep-th/0702176}.

\bibitem{Nilles:2011aj}
H.~P. Nilles, S.~Ramos-S{\'a}nchez, P.~K.~S. Vaudrevange, and A.~Wingerter,
  \emph{{The Orbifolder: A Tool to study the Low Energy Effective Theory of
  Heterotic Orbifolds}}, Comput. Phys. Commun. \textbf{183} (2012), 1363--1380,
  \texttt{arXiv:1110.5229} [hep-th].

\bibitem{Lebedev:2007hv}
O.~Lebedev, H.~P. Nilles, S.~Raby, S.~Ramos-S{\'a}nchez, M.~Ratz, P.~K.~S.
  Vaudrevange, and A.~Wingerter, \emph{The heterotic road to the {MSSM} with
  {R} parity}, Phys. Rev. \textbf{D77} (2007), 046013, \texttt{arXiv:0708.2691
  [hep-th]}.

\bibitem{Lebedev:2008un}
O.~Lebedev, H.~P. Nilles, S.~Ramos-S\'{a}nchez, M.~Ratz, and P.~K.~S.
  Vaudrevange, \emph{{Heterotic mini-landscape (II): completing the search for
  MSSM vacua in a $Z_6$ orbifold}}, Phys. Lett. \textbf{B668} (2008), 331--335,
  \texttt{arXiv:0807.4384} [hep-th].

\bibitem{Goodsell:2011wn}
M.~Goodsell, S.~Ramos-S{\'a}nchez, and A.~Ringwald, \emph{{Kinetic Mixing of
  U(1)s in Heterotic Orbifolds}}, JHEP \textbf{01} (2012), 021,
  \texttt{arXiv:1110.6901} [hep-th].

\bibitem{Carballo-Perez:2016ooy}
B.~Carballo-P{\'e}rez, E.~Peinado, and S.~Ramos-S{\'a}nchez,
  \emph{{$\Delta(54)$ flavor phenomenology and strings}}, JHEP \textbf{12}
  (2016), 131, \texttt{arXiv:1607.06812} [hep-ph].

\bibitem{Olguin-Trejo:2018wpw}
Y.~Olgu{\'i}n-Trejo, R.~P{\'e}rez-Mart{\'i}nez, and S.~Ramos-S{\'a}nchez,
  \emph{{Charting the flavor landscape of MSSM-like Abelian heterotic
  orbifolds}}, Phys. Rev. \textbf{D98} (2018), no.~10, 106020,
  \texttt{arXiv:1808.06622} [hep-th].

\bibitem{Olguin-Trejo:2019hxk}
Y.~Olgu{\'i}n-Trejo, O.~P{\'e}rez-Figueroa, R.~P{\'e}rez-Mart{\'i}nez, and
  S.~Ramos-S{\'a}nchez, \emph{{U(1)' coupling constant at low energies from
  heterotic orbifolds}}, Phys. Lett. B \textbf{795} (2019), 673--681,
  \texttt{arXiv:1901.10102} [hep-ph].

\bibitem{Mutter:2018sra}
A.~M\"utter, E.~Parr, and P.~K.~S. Vaudrevange, \emph{{Deep learning in the
  heterotic orbifold landscape}}, Nucl. Phys. B \textbf{940} (2019), 113--129,
  \texttt{arXiv:1811.05993} [hep-th].

\bibitem{Parr:2019bta}
E.~Parr and P.~K.~S. Vaudrevange, \emph{{Contrast data mining for the MSSM from
  strings}}, Nucl. Phys. \textbf{B952} (2020), 114922,
  \texttt{arXiv:1910.13473} [hep-th].

\bibitem{Parr:2020oar}
E.~Parr, P.~K.~S. Vaudrevange, and M.~Wimmer, \emph{{Predicting the orbifold
  origin of the MSSM}}, Fortsch. Phys. \textbf{68} (2020), no.~5, 2000032,
  \texttt{arXiv:2003.01732} [hep-th].

\bibitem{Escalante-Notario:2022fik}
E.~Escalante-Notario, I.~Portillo-Castillo, and S.~Ramos-S{\'a}nchez, \emph{{An
  autoencoder for heterotic orbifolds with arbitrary geometry}}, J. Phys. Comm.
  \textbf{8} (2024), no.~2, 025003, \texttt{arXiv:2212.00821} [hep-th].

\bibitem{Cervantes:2023wti}
E.~Cervantes, O.~P{\'e}rez-Figueroa, R.~P{\'e}rez-Mart{\'i}nez, and
  S.~Ramos-S{\'a}nchez, \emph{{Higgs-portal dark matter from nonsupersymmetric
  strings}}, Phys. Rev. D \textbf{107} (2023), no.~11, 115007,
  \texttt{arXiv:2302.08520} [hep-ph].

\bibitem{Rohm:1983aq}
R.~Rohm, \emph{{Spontaneous Supersymmetry Breaking in Supersymmetric String
  Theories}}, Nucl. Phys. B \textbf{237} (1984), 553--572.

\bibitem{Ibanez:1986tp}
L.~E. Ib{\'a}{\~n}ez, H.~P. Nilles, and F.~Quevedo, \emph{{Orbifolds and Wilson
  Lines}}, Phys. Lett. B \textbf{187} (1987), 25--32.

\bibitem{website:2025}
E.~Escalante-Notario, R.~P{\'e}rez-Mart{\'i}nez, S.~Ramos-S{\'a}nchez, and
  P.~K. Vaudrevange, \emph{The non-susy orbifolder}, 2025,
  {\texttt{http://stringpheno.fisica.unam.mx/nonSUSYorbifolder}}.

\bibitem{Green:1984sg}
M.~B. Green and J.~H. Schwarz, \emph{{Anomaly Cancellation in Supersymmetric
  $D=10$ Gauge Theory and Superstring Theory}}, Phys. Lett. \textbf{B149}
  (1984), 117--122.

\bibitem{website:2021}
R.~P{\'e}rez-Mart{\'i}nez, S.~Ramos-S{\'a}nchez, and P.~K. Vaudrevange,
  \emph{Non-supersymmetric orbifolds: model definitions and spectra}, 2021,
  {\texttt{http://stringpheno.fisica.unam.mx/nonsusy-orbifolds/}}.

\bibitem{Lauer:1989ax}
J.~Lauer, J.~Mas, and H.~P. Nilles, \emph{{Duality and the Role of
  Nonperturbative Effects on the World Sheet}}, Phys. Lett. \textbf{B226}
  (1989), 251--256.

\bibitem{Lauer:1990tm}
J.~Lauer, J.~Mas, and H.~P. Nilles, \emph{{Twisted sector representations of
  discrete background symmetries for two-dimensional orbifolds}}, Nucl. Phys.
  \textbf{B351} (1991), 353--424.

\bibitem{Kobayashi:2006wq}
T.~Kobayashi, H.~P. Nilles, F.~Pl{\"o}ger, S.~Raby, and M.~Ratz, \emph{{Stringy
  origin of non-Abelian discrete flavor symmetries}}, Nucl. Phys. B
  \textbf{768} (2007), 135--156, \texttt{hep-ph/0611020}.

\bibitem{Baur:2019iai}
A.~Baur, H.~P. Nilles, A.~Trautner, and P.~K.~S. Vaudrevange, \emph{{A String
  Theory of Flavor and $\mathcal{CP}$}}, Nucl. Phys. B \textbf{947} (2019),
  114737, \texttt{arXiv:1908.00805} [hep-th].

\bibitem{Nilles:2020gvu}
H.~P. Nilles, S.~Ramos-S\'anchez, and P.~K.~S. Vaudrevange, \emph{{Eclectic
  flavor scheme from ten-dimensional string theory - II. Detailed technical
  analysis}}, Nucl. Phys. B \textbf{966} (2021), 115367,
  \texttt{arXiv:2010.13798} [hep-th].

\bibitem{Baur:2021mtl}
A.~Baur, M.~Kade, H.~P. Nilles, S.~Ramos-S{\'a}nchez, and P.~K.~S. Vaudrevange,
  \emph{{Completing the eclectic flavor scheme of the $\mathbb Z_2$ orbifold}},
  JHEP \textbf{06} (2021), 110, \texttt{arXiv:2104.03981} [hep-th].

\bibitem{Baur:2024qzo}
A.~Baur, H.~P. Nilles, S.~Ramos-S{\'a}nchez, A.~Trautner, and P.~K.~S.
  Vaudrevange, \emph{{The eclectic flavor symmetries of
  $\mathbb{T}^2/\mathbb{Z}_K$ orbifolds}}, JHEP \textbf{09} (2024), 159,
  \texttt{arXiv:2405.20378} [hep-th].

\bibitem{Funakoshi:2025lxs}
S.~Funakoshi, Y.~Koga, and H.~Otsuka, \emph{{Classification of Modular
  Symmetries in Non-Supersymmetric Heterotic String theories}},  (2025),
  \texttt{arXiv:2503.23741} [hep-th].

\bibitem{Kobayashi:2011cw}
T.~Kobayashi, S.~L. Parameswaran, S.~Ramos-S{\'a}nchez, and I.~Zavala,
  \emph{{Revisiting Coupling Selection Rules in Heterotic Orbifold Models}},
  JHEP \textbf{05} (2012), 008, \texttt{arXiv:1107.2137} [hep-th], [Erratum:
  JHEP12,049(2012)].

\bibitem{Nilles:2013lda}
H.~P. Nilles, S.~Ramos-S{\'a}nchez, M.~Ratz, and P.~K.~S. Vaudrevange, \emph{{A
  note on discrete $R$ symmetries in $\Z{6}$-$\mathrm{II}$ orbifolds with
  Wilson lines}}, Phys. Lett. \textbf{B726} (2013), 876--881,
  \texttt{arXiv:1308.3435} [hep-th].

\bibitem{Dong:2025pah}
J.~Dong, T.~Kobayashi, R.~Nishida, S.~Nishimura, and H.~Otsuka, \emph{{Coupling
  Selection Rules in Heterotic Calabi-Yau Compactifications}},  (2025),
  \texttt{arXiv:2504.09773} [hep-th].

\bibitem{Acharya:2020hsc}
B.~S. Acharya, G.~Aldazabal, E.~Andr\'es, A.~Font, K.~Narain, and I.~G. Zadeh,
  \emph{{Stringy Tachyonic Instabilities of Non-Supersymmetric Ricci Flat
  Backgrounds}}, JHEP \textbf{04} (2021), 026, \texttt{arXiv:2010.02933}
  [hep-th].

\bibitem{Konopka:2012gy}
S.~J.~H. Konopka, \emph{{Non Abelian orbifold compactifications of the
  heterotic string}}, JHEP \textbf{07} (2013), 023, \texttt{arXiv:1210.5040}
  [hep-th].

\bibitem{Fischer:2013qza}
M.~Fischer, S.~Ramos-S{\'a}nchez, and P.~K.~S. Vaudrevange, \emph{{Heterotic
  non-Abelian orbifolds}}, JHEP \textbf{07} (2013), 080,
  \texttt{arXiv:1304.7742} [hep-th].

\bibitem{Narain:1986qm}
K.~S. Narain, M.~H. Sarmadi, and C.~Vafa, \emph{{Asymmetric Orbifolds}}, Nucl.
  Phys. B \textbf{288} (1987), 551.

\bibitem{Aldazabal:2025zht}
G.~Aldazabal, E.~Andr\'es, A.~Font, K.~Narain, and I.~G. Zadeh,
  \emph{{Asymmetric Orbifolds, Rank Reduction and Heterotic Islands}},  (2025),
  \texttt{arXiv:2501.17228} [hep-th].

\end{thebibliography}

\providecommand{\bysame}{\leavevmode\hbox to3em{\hrulefill}\thinspace}

\end{document}